\documentclass[pdftex,twocolumn,epjc3]{svjour3}          

\RequirePackage[T1]{fontenc}

\smartqed  

\RequirePackage{graphicx}
\RequirePackage{mathptmx}      
\RequirePackage{flushend}
\RequirePackage[numbers,sort&compress]{natbib} 
\RequirePackage[colorlinks,citecolor=blue,urlcolor=blue,linkcolor=blue]{hyperref}

\usepackage{amssymb}
\usepackage{mathtools}
\usepackage{amsfonts}
\usepackage{cancel}
\usepackage{cuted}
\usepackage{appendix}
\usepackage{subfig}
\usepackage{graphicx}
\usepackage[labelfont=bf]{caption}
\usepackage{hyperref}
\usepackage{float}
\usepackage{amsmath}
\usepackage{physics}
\usepackage{braket}
\usepackage{caption}
\usepackage{soul}

\usepackage{booktabs}

\usepackage[dvipsnames]{xcolor}
\journalname{Eur. Phys. J. C}

\definecolor{C3}{RGB}{214, 39, 40}
\definecolor{C0}{RGB}{31, 119, 180}
\definecolor{C2}{RGB}{44, 160, 44}

\definecolor{vir0}{RGB}{68, 1, 84}
\definecolor{vir1}{RGB}{33, 145, 140}
\definecolor{vir2}{RGB}{253, 231, 37}

\makeatletter
\renewcommand{\l@section}{\@dottedtocline{1}{0.5em}{3.5em}}
\makeatother

\renewcommand\appendixname{App.}

\begin{document}

\title{Optimal operation of cryogenic calorimeters through deep reinforcement learning}


\author{
  G.~Angloher\thanksref{addrMPI}\and
  S.~Banik\thanksref{addrHEPHY,addrAI}\and
  G.~Benato\thanksref{addrLNGS}\and
  A.~Bento\thanksref{addrMPI,addrCoimbra}\and 
  A.~Bertolini\thanksref{addrMPI}\and 
  R.~Breier\thanksref{addrBratislava}\and
  C.~Bucci\thanksref{addrLNGS}\and 
  J.~Burkhart\thanksref{addrHEPHY}
  L.~Canonica\thanksref{addrMPI}\and 
  A.~D'Addabbo\thanksref{addrLNGS}\and
  S.~Di~Lorenzo\thanksref{addrMPI}\and
  L.~Einfalt\thanksref{addrHEPHY,addrAI}\and
  A.~Erb\thanksref{addrTUM,addrWMI}\and
  F.~v.~Feilitzsch\thanksref{addrTUM}\and 
  S.~Fichtinger\thanksref{addrHEPHY}\and
  D.~Fuchs\thanksref{addrMPI}\and 
  A.~Garai\thanksref{addrMPI}\and 
  V.M.~Ghete\thanksref{addrHEPHY}\and
  P.~Gorla\thanksref{addrLNGS}\and
  P.V.~Guillaumon\thanksref{addrLNGS}\and
  S.~Gupta\thanksref{addrHEPHY}\and 
  D.~Hauff\thanksref{addrMPI}\and 
  M.~Ješkovsk\'y\thanksref{addrBratislava}\and
  J.~Jochum\thanksref{addrTUE}\and
  M.~Kaznacheeva\thanksref{addrTUM}\and
  A.~Kinast\thanksref{addrTUM}\and
  S.~Kuckuk\thanksref{addrTUE}\and
  H.~Kluck\thanksref{addrHEPHY}\and
  H.~Kraus\thanksref{addrOxford}\and 
  A.~Langenk\"amper\thanksref{addrMPI}\and 
  M.~Mancuso\thanksref{addrMPI}\and
  L.~Marini\thanksref{addrLNGS}\and
  B.~Mauri\thanksref{addrMPI}\and
  L.~Meyer\thanksref{addrTUE}\and
  V.~Mokina\thanksref{addrHEPHY}\and
  K.~Niedermayer\thanksref{addrHEPHY,addrAI} \and
  M.~Olmi\thanksref{addrLNGS}\and
  T.~Ortmann\thanksref{addrTUM}\and
  C.~Pagliarone\thanksref{addrLNGS,addrCASS}\and
  L.~Pattavina\thanksref{addrLNGS}\and
  F.~Petricca\thanksref{addrMPI}\and 
  W.~Potzel\thanksref{addrTUM}\and 
  P.~Povinec\thanksref{addrBratislava}\and
  F.~Pr\"obst\thanksref{addrMPI}\and
  F.~Pucci\thanksref{addrMPI}\and 
  F.~Reindl\thanksref{addrHEPHY,addrAI} \and
  J.~Rothe\thanksref{addrTUM}\and 
  K.~Sch\"affner\thanksref{addrMPI}\and 
  J.~Schieck\thanksref{addrHEPHY,addrAI}\and 
  S.~Sch\"onert\thanksref{addrTUM}\and 
  C.~Schwertner\thanksref{addrHEPHY,addrAI}\and
  M.~Stahlberg\thanksref{addrMPI}\and 
  L.~Stodolsky\thanksref{addrMPI}\and 
  C.~Strandhagen\thanksref{addrTUE}\and
  R.~Strauss\thanksref{addrTUM}\and
  I.~Usherov\thanksref{addrTUE}\and
  F.~Wagner\thanksref{addrHEPHY, cor}\and 
  V.~Wagner\thanksref{addrTUM}\and 
  M.~Willers\thanksref{addrTUM}\and 
  V.~Zema\thanksref{addrMPI}
  (CRESST Collaboration) \and
  C.~Heitzinger\thanksref{addrASC,addrCAIML} \and
  W.~Waltenberger\thanksref{addrHEPHY}
}

\institute
{Max-Planck-Institut f\"ur Physik, D-80805 M\"unchen, Germany \label{addrMPI} \and
Institut f\"ur Hochenergiephysik der \"Osterreichischen Akademie der Wissenschaften, A-1050 Wien, Austria\label{addrHEPHY} \and
Atominstitut, Technische Universit\"at Wien, A-1020 Wien, Austria \label{addrAI} \and
INFN, Laboratori Nazionali del Gran Sasso, I-67100 Assergi, Italy \label{addrLNGS} \and
Comenius University, Faculty of Mathematics, Physics and Informatics, 84248 Bratislava, Slovakia \label{addrBratislava} \and
Physik-Department, TUM School of Natural Sciences, Technische Universit\"at M\"unchen, D-85747 Garching, Germany \label{addrTUM} \and
Eberhard-Karls-Universit\"at T\"ubingen, D-72076 T\"ubingen, Germany \label{addrTUE} \and
Department of Physics, University of Oxford, Oxford OX1 3RH, United Kingdom \label{addrOxford} \and
Institute of Information Systems Engineering, Technische Universit\"at Wien, A-1040 Wien, Austria \label{addrASC} \and
Center for Artificial Intelligence and Machine Learning, Technische Universit\"at Wien, A-1040 Wien, Austria \label{addrCAIML} \and
also at: LIBPhys-UC, Departamento de Fisica, Universidade de Coimbra, P3004 516 Coimbra, Portugal \label{addrCoimbra} \and
also at: Walther-Mei\ss ner-Institut f\"ur Tieftemperaturforschung, D-85748 Garching, Germany \label{addrWMI} \and
also at: GSSI-Gran Sasso Science Institute, I-67100 L'Aquila, Italy \label{addrGSSI} \and
also at: Dipartimento di Ingegneria Civile e Meccanica, Universitá degli Studi di Cassino e del Lazio Meridionale, I-03043 Cassino, Italy\label{addrCASS}
}


\thankstext{cor}{e-mail: felix.wagner@oeaw.ac.at}



\maketitle

\begin{abstract}
Cryogenic phonon detectors with transition-edge sensors achieve the best sensitivity to sub-GeV/c$^2$ dark matter interactions with nuclei in current direct detection experiments. In such devices, the temperature of the thermometer and the bias current in its readout circuit need careful optimization to achieve optimal detector performance. This task is not trivial and is typically done manually by an expert. In our work, we automated the procedure with reinforcement learning in two settings. First, we trained on a simulation of the response of three CRESST detectors used as a virtual reinforcement learning environment. Second, we trained live on the same detectors operated in the CRESST underground setup. In both cases, we were able to optimize a standard detector as fast and with comparable results as human experts. Our method enables the tuning of large-scale cryogenic detector setups with minimal manual interventions.
\end{abstract}

\tableofcontents

\section{Introduction}
\label{sec:intro}

\begin{figure*}[!t]
\centering
\includegraphics[width=\textwidth]{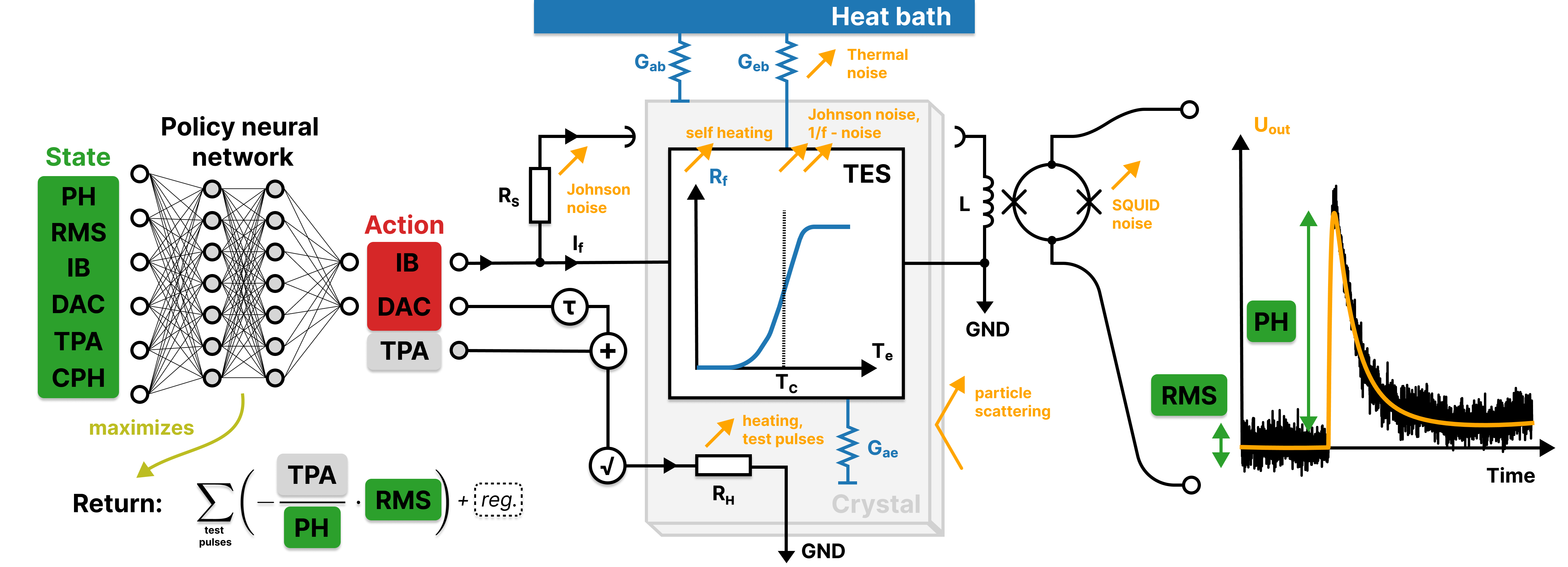}
\caption{Schematic drawing of the detector environment. The circuits are schematical visualizations and not complete electrical and thermal circuits. (center) The detector can be described as an electrothermal system, where the readout and heater electronics and the temperatures in the crystal and sensor interact with each other. Visualizations of the thermal system are in blue and of the electrical system in black. The readout circuit of the TES (central in the figure) and the heater circuit (lower center) are electrically separated. (right) The recorded observable from particle recoils is a pulse-shaped voltage signal (orange) in superposition with sensor noise (black). Features to quantify the quality of the detector response can be extracted from the pulse shape ($PH$, $RMS$). (left) A policy neural network is trained with RL to choose optimal control parameters based on the state of the system. Optimal control parameters maximize the return, a target function that is closely related to the SNR. A maximal return realizes a trade-off between low noise amplitude, linear range of the detector, and stable measurement conditions. See text in Sec.~\ref{sec:setup} and \ref{sec:Analysis} for details. }
\label{fig:promo_schematic}
\end{figure*}

The physics goals of several types of particle physics experiments require the sensitive measurement of low-energy particle recoils, e.g.~direct detection dark matter (DM) searches and coherent neutrino-nucleus scattering. One successful detector concept is that of cryogenic phonon detectors with transition-edge sensors (TES), used by the CRESST experiment to reach the currently best sensitivity to sub-GeV/c$^2$ DM-nucleus interactions \cite{PhysRevD.100.102002}. These consist of a crystal and an attached TES, acting as a thermometer, where both are cooled to $\approx$~10~mK. A particle recoil in the crystal produces a temperature increase in the crystal and the TES, leading to a measurable signal for recoils with energies as low as 10~eV \cite{cresstcollaboration2022results}. Reaching the best sensitivity requires a careful setup of the detectors, effectively optimizing the bias current applied to the TES and the temperature of the system, which is controlled by the current applied to a heating resistor attached to the crystal. 

The optimization of these two parameters is typically performed by an expert and by hand. After each change to the control parameters, the system needs to reach its new equilibrium state before the sensitivity can be tested. Therefore, the optimization can be lengthy, taking up to several hours. The objectives of future physics experiments, e.g.~the planned CRESST upgrade \cite{Billard_2022}, require the simultaneous operation of several tens, up to hundreds, of detector modules. Automating and parallelizing the optimization is necessary to achieve this objective and stay within reasonable bounds of the required manual workload. The cost of cryostat operation is generally high, and faster optimization leads to a higher detector live time. 

The optimal parameter settings vary between similarly manufactured detectors due to fluctuations in the properties of the TES, crystal, and thermal couplings, and the true values are often unknown or have large uncertainties. Observed optimal values from identically produced detectors show significant variation. Therefore, a-priori predictions are not useful.
The detector's sensitivity is not uniquely determined by the setting of the control parameters but can, due to the Joule heating of the TES by its bias current, also depend on an internal state of the TES, namely the fact whether it was in a superconducting state before the parameters were set. This internal state can cause a hysteresis-like effect.
Simple approaches to optimizing the control parameters, such as a grid search or an educated guess, cannot be optimal in terms of an optimization time to performance trade-off.

In practice, a combination is often used: choosing a small set of educated guesses for bias currents and recording a one-dimensional sweep of the heating resistor, starting from warm to cold. The choice of control parameter configuration is then based on the amplitude of the detector response. While this approach can work, it has two weaknesses: a) it does not systematically account for different noise conditions, and b) the sweeps spend an unnecessarily large amount of measurement time in regions of the control parameter space that are unlikely to be optimal, based on the already assembled knowledge from previous observations. While a standard gradient-based or Bayesian optimization approach could improve b) it would also rely on a hard-coded mechanism to approach the desired point in parameter space or risk performing measurements with an unknown internal state of the superconductor.

The previous arguments make it clear that the solution to the optimization problem is not only the set of optimal parameters but also the awareness of an allowed way to approach them, i.e.~a sequence. We formulate the problem in the framework of reinforcement learning (RL), a general method to find optimal policies for control problems in discrete time, extensively described e.g.~in Ref.~\cite{sutton2018reinforcement}. 

In RL, we model the problem as the time-ordered interaction of an agent with an environment. Based on a learned policy, the agent takes actions that depend on its latest observation of the environment. The environment returns a new observation and a reward for each given action. The agent's objective is the maximization of returns, which are the sum of rewards over time. The estimated future returns for a given action-observation combination are called values. They are learned jointly with the policy. RL agents can adjust their actions to the environment's state, which gives them an advantage compared to state-independent optimizers. We exploit this advantage to handle the hysteresis and time-dependent effects of the detector. 
The optimization of control parameters with RL does not require manual interactions. Therefore, we can fully parallelize the procedure for all operated detectors, contrary to procedures that human experts must manually carry out. 

RL has already been used to optimize control settings in physics in Refs.~\cite{PhysRevAccelBeams.24.104601,PhysRevAccelBeams.23.124801,Velotti_2023} for particle beams, in Ref.~\cite{degrave_magnetic_2022} for nuclear fusion reactors and in Ref.~\cite{Nautrup_2019} for superconducting quantum bits. 

The state-of-the-art RL algorithms for finding optimal policies in an environment where the actions and observations are continuous values include Actor-Critic (AC) methods, where both policy and value function are approximated with neural networks. The Soft AC algorithm (SAC) has been shown to perform well in real-world applications, e.g.~in robotics \cite{haarnoja2019soft}. RL methods are typically associated with a low sampling efficiency, i.e.~they require many interactions between the agent and environment to discover an optimal policy. However, dedicated RL algorithms with reasonably high sampling efficiency exist. 

We provide proof-of-principle of our method in a virtual environment, modeled after three CRESST detectors operated between 2020 and 2023 and used for a DM search in Ref.~\cite{PhysRevD.106.092008}. Furthermore, we optimize these three detectors live and directly in the CRESST setup by interfacing the experiment control software with our RL agent. 

We present the following in this work:
\begin{itemize}
\item The general setup of cryogenic detectors with TES is introduced in Sec.~\ref{sec:DetectorIntro}.
\item We define a reward function that encodes the goal of maximizing the sensitivity of the detector based on observable detector response parameters and explain the approach of optimizing it with SAC agents in Sec.~\ref{sec:Analysis}. 
\item We train a SAC agent to optimize a cryogenic detector in our virtual environment in Sec.~\ref{sec:virtual}. The virtual environment is based on our model for the cryogenic detector response and noise contributions, described in Sec.~\ref{sec:setup}.
\item We train and operate on the CRESST setup live in Sec.~\ref{sec:live}.
\end{itemize}

\section{Cryogenic detectors with transition-edge sensors}
\label{sec:DetectorIntro}

\textit{The sensitivity of cryogenic detectors with TES depends on a delicate optimization of their heating and bias current, which determine the signal and noise conditions. The observed data from particle recoils are pulse-shaped voltage traces, and heater pulses can be used to monitor the detector response.}

A cryogenic detector consists of a monocrystalline target onto which a superconducting film, the TES, is evaporated or sputtered. The system is cooled in a dilution refrigerator below the transition temperature of the superconducting film. A heating resistor $R_H$, close to but electronically isolated from the film on the crystal, is then used to fine-tune the temperature of the film, such that it is operated in the transition between its normal and superconducting state. In this configuration, a small temperature increase causes a large increase in the resistance of the film, which can be measured when read out with a low-noise circuit. The situation is schematically depicted in Fig.~\ref{fig:promo_schematic} (center). The temperature increase caused by a particle recoil in the target is observed to have a pulse-like shape (see Fig.~\ref{fig:promo_schematic}, right). A model to describe the pulse shape is introduced in Sec.~\ref{sec:setup}. When the TES is operated in an approximately linear region of its transition curve, pulses are observed with a height directly proportional to the temperature increase in the film. When the temperature approaches the normal conducting region, the transition curve flattens. Therefore, the pulse from a high-energy particle recoil will be observed with a significantly flattened peak and possibly further distortions. 

The OP within the superconducting transition is the stable temperature-current-resistance combination to which the TES is heated while no pulses are observed. We use the term operating point synonymously for a combination of the control parameters $DAC$ and $IB$. The heating is mostly governed by the constant part of the current applied to the heating resistor. In our setup, a digital-analog converted ($DAC$) value ranging between 0 and 10 V is the controlling quantity for this constant current component. The heating resistor is also used to inject heater pulses to monitor the detector response. Two types of such pulses are injected: test pulses with test pulse amplitude ($TPA$), which is also a value between 0 and 10, are used to monitor the linearity of the detector response and the signal-to-noise ratio (SNR), while control pulses with maximal amplitude are used to drive the detector into the saturated region of its response, and ideally out of the superconducting transition, such that the OP can be inferred from their pulse height. The test pulses are designed to have a similar shape as particle pulses. The electrical power input of test and control pulses are exponentially decaying pulses on a short time scale, imitating the heat produced by thermalizing phonons by a fast component of the heater current. The total heater current is 
$$\propto \sqrt{DAC + TPA \cdot \mathcal{P}(t)},$$
\noindent where $\mathcal{P}(t)$ is the template of the injected heater pulse, resulting in two independent and linear controls for the electrical power input.
More details can be found in the mathematical model in Sec.~\ref{sec:setup}




The second control parameter for the detector system is the bias current $IB$ applied to the TES. The applied current changes the transition temperature and shape of the transition curve of the superconductor. The response is different in each sample. The applied current also causes Joule heating (self-heating) on the superconductor. The electrical power dissipated in the TES depends on its momentary resistance and, therefore, interacts with the pulse shape of temperature increases, causing electro-thermal feedback (ETF). 
Since both the DAC and the IB parameters change the temperature of the TES, a certain degree of degeneracy is expected between the effects of the control parameters. 
Both parameters have an effect on the readout noise independently. However, this impact is more complex and in more detail discussed in Sec.~\ref{sec:setup}. Generally, a higher bias current can emphasize the pulse height and intrinsic sensor noise over external noise sources, which in most cases positively impacts the overall SNR. An optimal combination of the $DAC$ and $IB$ has to be found by solving an optimization problem, which will be discussed in more detail in Sec.~\ref{sec:Analysis}.

\section{Optimizing the sensitivity}
\label{sec:Analysis}

\textit{The detector's energy threshold can be estimated from the pulse height and noise conditions of injected heater pulses, and its minimization can be framed as a numerical optimization problem. The target function is closely related to the signal-to-noise ratio, and minimizing it for a set of heater pulses results in a trade-off between low noise, linear range, and stable measurement conditions. We propose to solve the optimization problem with reinforcement learning. The general framework and the specific algorithm, the Soft Actor-Critic algorithm, that we used for our experiments are introduced in this section.}

The low energy threshold $E_{th}$ of TES detectors is the crucial parameter, making them the ideal choice for many physics cases, such as light DM search. 
We define a low energy threshold as the optimization objective for the control parameters $DAC$ and $IB$. The optimization problem can be written in terms of easily accessible observables for a sequence of $N$ injected test pulses. A detailed derivation, contained in \ref{app:reward}, leads to the target function: 

\begin{equation}
    \underset{DAC,\ IB}{\arg\max}\left( - \sum_{i=1}^N \frac{TPA_i}{PH_i} RMS_i\right), \label{eq:raw_target}
\end{equation}

\noindent where $TPA_i$, $PH_i$ and $RMS_i$ are the $TPA$, pulse height, and root-mean-square ($RMS$) value of the noise in the pre-trigger region of the $i$'th test pulse, respectively. Intuitively, Eq.~(\ref{eq:raw_target}) can be interpreted as the negative, inverse SNR within linear regions of detector response. However, if heater pulses are high enough to reach the saturated region of detector response, their pulse height additionally decreases. In this regime, the optimization objective is a trade-off between best noise conditions and a long linear region of detector response. The control values $DAC$ and $IB$ that optimize Eq.(~\ref{eq:raw_target}) realize a trade-off between lowest noise conditions, avoidance of saturation effects (linear range), and stable conditions, where the weighting between these three targets is determined by the choice of $TPA$ values and the environmental conditions.
For most of our experiments, we are mostly interested in achieving the best possible SNR for small pulses. We can, therefore, additionally weight the test pulses by the inverse of their $TPA$ value to achieve an additional emphasis on small injected pulses:

\begin{equation}
    \underset{DAC,\ IB}{\arg\max}\left( - \sum_{i=1}^N \frac{1}{PH_i} RMS_i\right). \label{eq:adapted_target}
\end{equation}

\noindent All quantities appearing in the equation above -- except for $TPA$ -- depend on $DAC$ and $IB$. The adapted target Eq.~(\ref{eq:adapted_target}) has several convenient properties:

\begin{enumerate}
    \item The $TPA$ of the injected pulse is not explicitly contained anymore, and the derived function would also work as a target to optimize a detector with triggered pulses from a particle source. 
    \item The function can be evaluated on an event-by-event basis by measuring the noise $RMS$ in the pre-trigger region of a record window containing a pulse and by taking the maximum value in the record window as pulse height $PH$. 
    \item The target function is always negative, and its upper bound is the value zero, which cannot be attained. Furthermore, we can restrict the function to values larger than minus one, as other values can only occur when a record is corrupted by artifacts in the pre-trigger region, e.g.~by negative voltage spikes. 
\end{enumerate}

\noindent Evaluating Eq.~(\ref{eq:adapted_target}) for individual events leads to fluctuating values due to the natural randomness of the sensor noise, but it is suitable to be used as a target function in a time-dependent optimization problem. We discuss in the following the RL framework, which we use to solve the above optimization problem.

\subsection{Reinforcement learning}

\begin{figure}[!t]
\centering
\includegraphics[width=.7\linewidth]{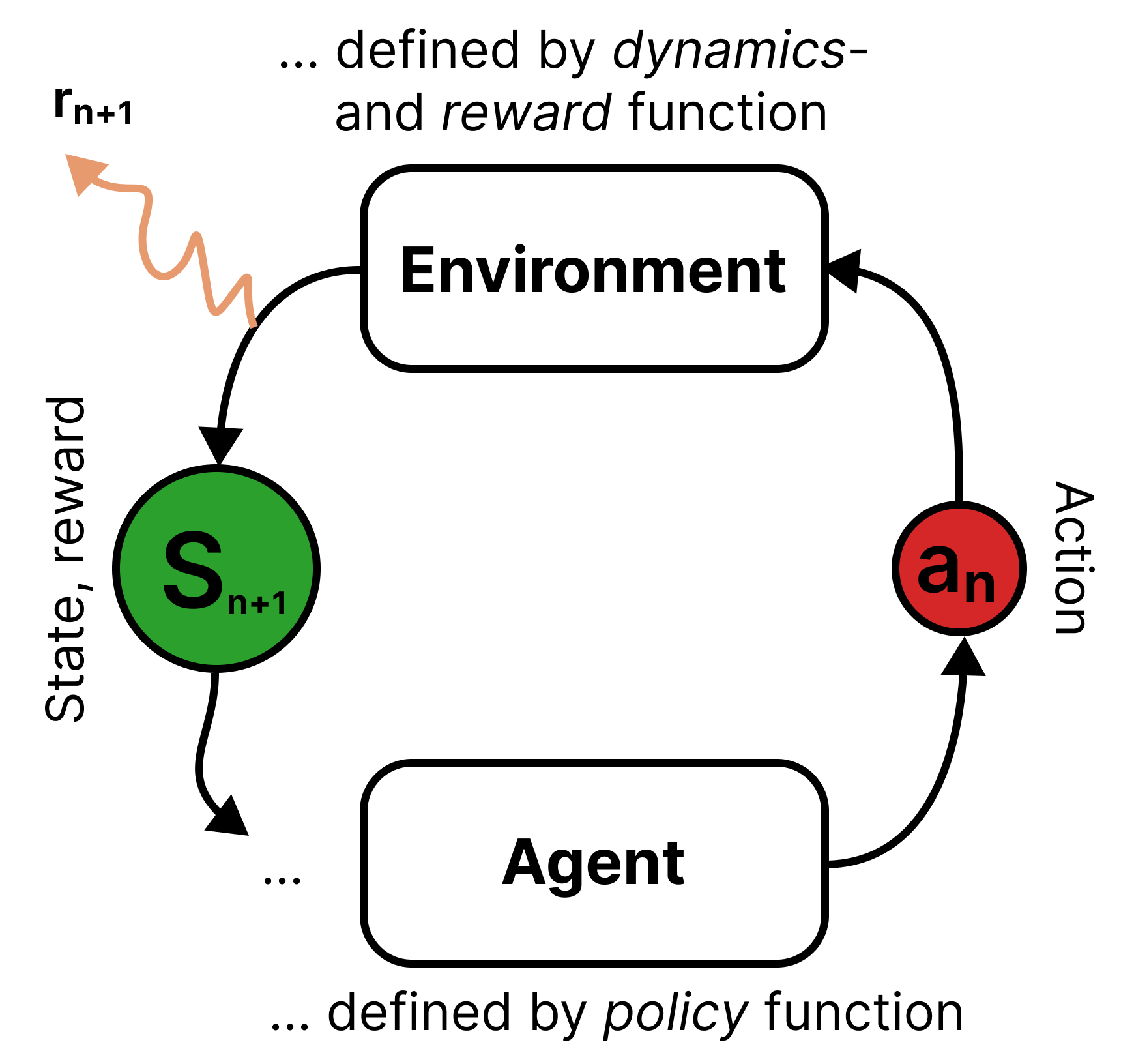}
\caption{The mechanics of RL: an agent follows a policy function to interact with an environment. The environment, defined by its dynamics and reward function, responds to the agent's actions with a reward and observable state. Figure adapted from Ref.~\cite{sutton2018reinforcement}.}
\label{fig:rl_cycle}
\end{figure}

RL is a general framework for optimizing time-dependent control problems. We provide here a short summary of the necessary vocabulary and definitions. For an extensive introduction, we refer to Ref.~\cite{sutton2018reinforcement}. The original formulation uses the framework of Markov decision processes (MDPs), which are defined as a 4-tuple of a state space $\mathcal{S}$, an action space $\mathcal{A}$, a dynamics function 

$$ p\colon\quad (A, S) \mapsto\left\{\text {probabilities for } S^{\prime}\right\},\hspace{0.3cm} A \in \mathcal{A},\hspace{0.3cm} S, S^{\prime} \in \mathcal{S}, $$

\noindent which defines state transitions for action-state pairs, and a reward function

$$ r\colon\quad\left(S, A, S^{\prime}\right) \mapsto R \in \mathbb{R} $$

\noindent that assigns scalars to state transitions. The dynamics and reward function jointly define an environment with an observable state that can be interacted with through actions. The goal of RL is to find a policy function 

$$ \pi\colon\quad S \mapsto \pi(A \mid S)=\{\text {probabilities for } A \text { given state } S\} $$

\noindent that maximizes the return $R$, the sum of collected rewards over time. The policy function is thought of as an agent that interacts with the environment and learns through experience. This framework is schematically summarized in Fig.~\ref{fig:rl_cycle}. Dynamics and reward function are not directly observable for the agent but must be approximated from experience. The definition of an MDP automatically satisfies the Markov property, i.e.~$p$ and $r$ only depend on the current state and action and not on prior history. For many practical applications, the state of the environment is not fully observable, and the Markov property is not necessarily fulfilled with the observable state. 

For framing detector operation as a reinforcement learning problem, we define the state and action spaces:

\begin{align*}
\mathcal{S} &\coloneqq \{ PH, RMS, IB, DAC, TPA, CPH \}, \\
\mathcal{A} &\coloneqq \{ DAC, IB \},    
\end{align*}

\noindent where $PH$ and $RMS$ are the maximum and RMS of the pretrigger region of an injected test pulse, $IB$ and $DAC$ are the set control parameters at the time of recording the pulse, and $TPA$ is the amplitude of the injected voltage pulse. Additionally, after every test pulse with defined $TPA$, we choose to inject a control pulse and include its pulse height $CPH$ in the state. The values of $PH$ and $RMS$ scale with the applied bias current. To reduce the complexity of the state space, we divide them by $IB$ and only allow for positive bias currents. We normalize all action and state values such that they are continuous values within the range $(-1,1)$. We use Eq.~(\ref{eq:adapted_target}) as our reward function for all experiments conducted in Sec.~\ref{sec:virtual}. In Sec.~\ref{sec:live} we use both Eq.~(\ref{eq:adapted_target}) and Eq.~(\ref{eq:adapted_target}). 

\subsection{The Soft Actor-Critic algorithm}

The SAC algorithm showed good performance in Ref.~\cite{haarnoja2019soft} in a real-world robotics task with continuous state and action spaces, and we therefore chose it for our application.

AC algorithms use -- additionally to the policy function $\pi$, in this context also called an actor -- a value function, or critic
$$q\colon\quad(S,A) \mapsto Q \in \mathbb{R},  $$
\noindent which maps action state pairs to estimates of the future return. For the function approximators of both policy and critic we use neural networks $\pi_\phi$, $q_\theta$ and train their weights $\phi$, $\theta$ with gradient descent. For the policy function, we parameterize with the outputs of the neural network a Gaussian function with the dimensionality of the action space to obtain an explicit conditional probability distribution for actions in a given state. The collected experience is stored in an experience replay buffer from where state transitions  $(S,A,R,S^\prime)$ are sampled as training data. 

The critic is trained to minimize the soft Bellman residual:  
$$ J_{q}( \theta ) \propto \left( q_{\theta }( S,A) -\left( R+\gamma \left( q_{\overline{\theta }}( S',a') -\alpha \ln \pi _{\phi }( a'|S')\right)\right)\right)^{2},$$
\noindent with $a' \sim \pi_\phi(\cdot|S')$, and where the term with the hyperparameter $\alpha$ as coefficient is designed to encourage exploration. $\alpha$ is called the temperature. The discount factor $\gamma$ is introduced for the numerical stability of long trajectories and is not to be confused with the similarly named factor used for the reward derivation in \ref{app:reward}. The weights $\overline{\theta}$ of the target critic are discussed later in this section.

The target function minimized by the policy quantifies the Pareto-optimum between exploration and exploitation:
$$ J_{\pi }( \phi ) \propto \alpha \ln \pi _{\phi }( a|S) -q_{\theta }( S,a). $$
\noindent with $a \sim \pi_\phi(\cdot|S)$. There are several technical details of this algorithm that stabilize the training procedure and the exploration versus exploitation trade-off:
\begin{itemize}
    \item Two critics are trained simultaneously, and the minimum of their outputs is used for inference.
    \item The loss function for the critics uses predicted values by their target critics. These are versions of the neural networks with weights $\overline{\theta}$ that are obtained by exponentially smoothing the critic weights $\theta$.
    \item The value of $\alpha$ is automatically adjusted jointly with the gradient steps done for the neural networks. The objective of $\alpha$ is to realize a pre-defined target entropy of the policy function, i.e.~a certain width of the Gaussian. In contrast to the default algorithm introduced in \cite{haarnoja2019soft}, we adjust the target entropy during the training such that the policy can converge towards smaller features of the parameter space when training progresses. A motivation for this feature is described in \ref{app:entropy}. 
\end{itemize}
SAC is an off-policy algorithm, i.e.~the policy function that is learned during training is not necessarily the policy that was used to collect the experience. The fact that data collection and training are two independent processes is useful for practical applications and is exploited in Sec.~\ref{sec:live}.

\section{Operation in a virtual environment} \label{sec:virtual}

\textit{We simulated the response of three detectors currently operated in the CRESST setup and wrapped it as an OpenAI Gym reinforcement learning environment. Within this virtual environment, we tested the capability of Soft-Actor Critic agents to perform the optimization of the control parameters. In total, we trained 315 agents on variations of the simulated detectors and hyperparameters of the algorithm. We showed that in the simulation, we can reach the performance of a human expert, both in terms of optimality of the found operation points and in terms of necessary equivalent measurement time for the optimization.}

\subsection{Modeling the detector response and noise}
\label{sec:setup}

The detector's response to particle recoils and other energy depositions depends on the thermal properties of the absorber crystal and TES and the electrical properties of the readout circuit. We can calculate a simplified model of the expected detector response by independently modeling the thermal and electrical circuits involved with ordinary differential equations (ODEs) and solving them jointly as a coupled ODE system. These response calculations were performed analytically for the coupled thermal system of the absorber and TES in Ref.~\cite{probst_model_1995} in a small-signal approximation. Analytical calculations for the coupled system of an isolated TES's thermal and electrical response were studied in Ref.~\cite{irwin_transition-edge_2005}. 

The interacting components are schematically drawn in Fig.~\ref{fig:promo_schematic}. The crystal is symbolized by the grey block in the center, the TES by the white rectangle enclosing a sketched graph of the temperature-dependent resistance of the superconducting film $R_f(T)$ that drops sharply around the superconductors transition temperature $T_c$. The thermal circuit connects the temperature of the heat bath $T_b$ with the phonon temperatures of the crystal $T_a$ and the electron temperature of the TES $T_e$. The heat flow in the system is determined by the thermal connectivity between the components $G_{ae}$, their individual links to the heat bath $G_{eb}$ and $G_{ab}$, and the heat capacities of the absorber $C_a$ and TES $C_e$. The heat capacity of the TES increases below its superconducting transition by a factor of 2.43, see e.g.~Ref.~\cite{tinkham2004introduction}. The TES is operated in a parallel electric circuit with a shunt resistor $R_s$ and a readout coil with inductivity $L$. The circuit is biased with a current $I_b$ from a current source. We neglect the temperature dependence of all properties other than the TES resistance and heat capacity, which provides us with a tractable model for a small temperature range near the critical temperature. The electrical and thermal equations for the state variables are written in Eq.~(\ref{eq:odes}). They are coupled through the TES temperature. The system's state variables are the absorber and TES temperatures and the current through the TES branch of the readout circuit $I_f$. They are all time-dependent variables. However, we omit writing their time dependence explicitly for better readability:

\begin{align}
C_{e} (T_e) \frac{d T_{e}}{d t}+\left(T_{e}-T_{a}\right) G_{e a}+\left(T_{e}-T_{b}\right) G_{e b}&=P_{e}(t), \nonumber\\ 
C_{a} \frac{d T_{a}}{d t}+\left(T_{a}-T_{e}\right) G_{e a}+\left(T_{a}-T_{b}\right) G_{a b}&=P_{a}(t), \label{eq:odes}\\
L \frac{dI_f}{dt} + R_s I_b - (R_f(T_e) + R_s)I_f &= 0. \nonumber
\end{align}

\noindent The system responds to power inputs in the absorber $P_a$ and thermometer $P_e$, which are introduced by deposited energy $\Delta E$ e.g. from particle recoils in the crystal, heater pulses, or the constantly applied heating. 

We model injected heater pulses with a given $TPA$ as an exponential decay, we call their decay time $\tau_{TP}$. The controlling values for the heating current are summed and square rooted, such that they linearly control the power inputs. A particle recoil produces an initial, distinct population of athermal phonons of which a share $\epsilon$ thermalizes on a time scale $\tau_n$ in the TES and a share $(1 - \epsilon)$ in the absorber, mostly by surface scattering. Assuming an exponential time scale for thermalization is equivalent to assuming a monochromatic athermal phonon population, which is a sensible approximation for our purposes. 

Additional heat input in the system is due to the Joule heating by the bias current of the TES, with power $R_f I_f^2$. This contribution is crucial, as it strongly influences the TES temperature and introduces an internal state in the system. 
Multiple stable equilibrium states can exist with high and low bias heating. Which state the TES occupies depends on the past history of the system.
The heat inputs are summarized in Eq.~(\ref{eq:tes_heat}) and (\ref{eq:absorber_heat}), where we introduced factors $\delta$ and $\delta_H$ to distribute the power inputs from the constant heating and test pulses between the TES and absorber. These factors absorb the spatial dependence of the temperatures, which may be non-homogeneous across the system's geometry for the locally induced power from the heating resistor. Furthermore, they absorb the potentially different energy distributions of produced phonons in the constant heating and the fast heater pulses. The factors $\epsilon$, $\delta$, and $\delta_H$ are all values between zero and one. The power input from athermal phonon thermalization is in good approximation uniform across the geometry of the components, as they spread ballistically across the system on a much shorter time scale than they thermalize. 

Geometric effects of the temperature distribution in the TES were studied in Ref.~\cite{probst_model_1995}, where it was shown that such effects could be absorbed in the thermal parameters of the system. The evolution of the power inputs for a pulse onset at $t = t_0$ can be written as follows:

\begin{align}
P_{e}(t) &= P_{e,\text{pulse}}(t) + R_f I_f^2 + \delta_H \frac{DAC}{10} R_H I_H^2, \label{eq:tes_heat} \\
P_{a}(t) &= P_{a,\text{pulse}}(t) + (1 - \delta_H) \frac{DAC}{10} R_H I_H^2. \label{eq:absorber_heat}
\end{align}

\noindent with 

\begin{align}
    P_{e,\text{pulse}}(t) &= \theta(t-t_0) \epsilon \frac{\Delta E}{\tau_n} \exp(\frac{t-t_0}{\tau_n}), \\
    P_{a,\text{pulse}}(t) &= \theta(t-t_0) (1 - \epsilon) \frac{\Delta E}{\tau_n} \exp(\frac{t-t_0}{\tau_n}),
\end{align}

\noindent for particle pulses and 

\begin{align}
    P_{e,\text{pulse}}(t) &= \theta(t-t_0) \delta \frac{TPA}{10} \exp(\frac{t-t_0}{\tau_{TP}}) \beta R_H I_H^2, \\
    P_{a,\text{pulse}}(t) &= \theta(t-t_0) (1 - \delta) \frac{TPA}{10} \exp(\frac{t-t_0}{\tau_{TP}}) \beta R_H I_H^2,
\end{align}

\begin{figure*}[!t]
\centering
\includegraphics[width=\textwidth]{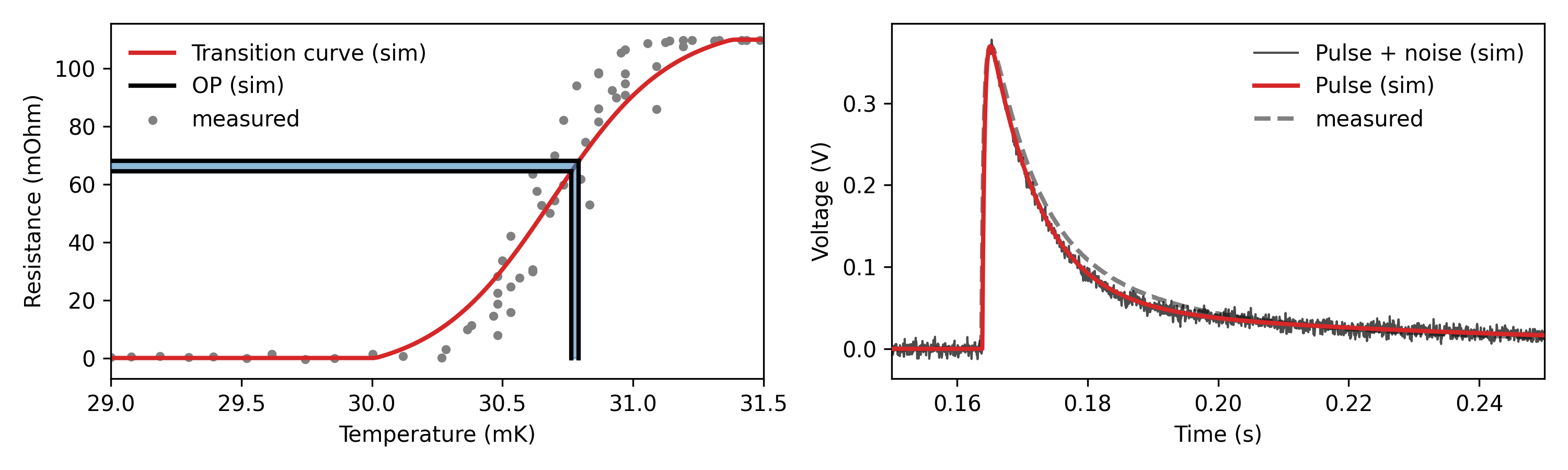} \quad 
\includegraphics[width=\textwidth]{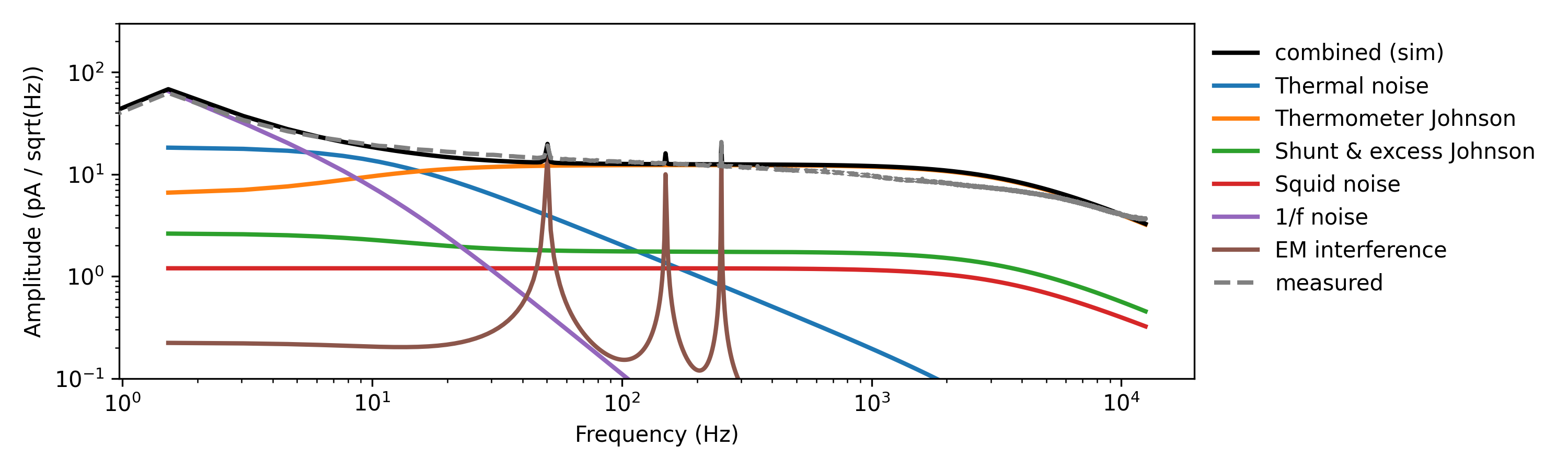} \quad 
\caption{Simulation and measurement of a 5.95 keV X-ray event induced by a calibration source in the Li1P detector. (upper left) The OP (black/blue lines) within the simulated transition curve of the TES (light red line). A measurement of the transition curve is shown for comparison (grey dots). (upper right) The voltage pulse induced in the simulated SQUID amplifier without noise (red) and overlayed with noise generated from the simulated NPS (black). A measured voltage pulse is shown for comparison (grey dashed). (lower part) The simulated NPS (black) has individual noise contributions (colored). The 1/f, excess Johnson, and EM interference noise components were adjusted to fit the measured NPS (grey dashed). }
\label{fig:pulse_noise}
\end{figure*}

\noindent for test pulses. The detector response is distorted by electrical and thermal noise produced in the system, leading to a finite energy resolution. The contributions to the observed noise power spectrum (NPS) can be modeled as stochastic fluctuations of the right-hand side of Eq.~(\ref{eq:odes}). The major noise contributions that we include in our model are: 

\begin{itemize}
    \item The thermal noise, or phonon noise, $\Delta I_{th}$. This noise arises from natural thermal fluctuations along the thermal coupling of the thermometer and bath, and its contribution is often sub-dominant.
    \item The Johnson noise of the TES $\Delta I_{Jf}$ and the shunt resistor $\Delta I_{Js}$. This noise comes from fluctuations in the movement of the electrons through the resistors that typically dominate the NPS for high frequencies. Excess electrical noise was observed in experiments and described in the literature (e.g.~Ref.~\cite{irwin_transition-edge_2005}) and can originate from other electrical components, setup uncertainties, or in the TES. We absorb such excess electrical noise by scaling $\Delta I_{Js}$ accordingly.
    \item The noise introduced by the superconducting quantum interference device (SQUID) amplifier $\Delta I_{sq}$, which measures the magnetic field introduced by $L$, i.e.~the final signal that is digitized and recorded. Its contribution is determined by $i_{sq}$, a constant value and property of the used SQUID system.
    \item The 1/f noise $\Delta I_1/f$, also called flicker noise. This noise appears across all TES and other devices, and its origin is not fully clarified. It, therefore, cannot be predicted precisely but depends on an empirical scale factor $\Delta R_{f,\text{flicker}}/R_{f0}$. Ref.~\cite{galeazzi_microcalorimeter_2003} proposes a connection of this noise with resistance fluctuations of the TES. It dominates the NPS for low frequencies and is the most harmful noise contribution.
    \item Several characteristic peaks in the NPS are introduced by the power supply voltage at 50 Hz and its harmonics.
\end{itemize}

\noindent Other known noise contributions exist but were omitted because they are sub-dominant or difficult to model. This includes internal fluctuation noise, burst noise, and any noise sources that would arise in the absorber crystal. 

To acquire useful descriptions of the detector response and noise, we solve Eq.~(\ref{eq:odes}) on two temperature scales independently: the macroscopic scale of observable, individual energy depositions from heating, particle recoils, or test pulses, and the microscopic scale of thermal and electrical fluctuations. We assume that these scales do not interact with each other.  

On a macroscopic scale, we solve the ODEs Eq.~(\ref{eq:odes}) numerically, such that we can include the non-linear dependencies of $C_e$ and $R_f$. We use SciPy’s \textit{odeint} method, a wrapper of the FORTRAN LSODA solver that is especially suitable for solving stiff ODE systems \cite{2020SciPy-NMeth}, to calculate the observed pulse shape. 
On the microscopic scale, the small signal approximation is very well satisfied, and we use it to derive explicit formulas for the observed NPS in a given OP, described in \ref{app:noise}. We use the method described in Ref.~\cite{carrettoni_generation_2010} to generate colored noise traces with the calculated NPS. Once pulse shape and noise in a given OP and for defined $P_e$ and $P_a$ are calculated, we superpose them and translate them with the known SQUID settings to a voltage trace that would be observed in a real-world setup.

The detector response can additionally depend on the trajectory through which an OP is approached. To include this large-scale time dependency, we start trajectories after resetting the virtual environment from an edge of the parameter space and solve the ODE system continuously with large mesh grid sizes for intervals without energy depositions and small ones for intervals where signals are simulated.

For the tests reported in Sec.~\ref{sec:virtual}, we adjust all parameters of our simulation to resemble the detector response and noise of three detectors currently operated in the CRESST experiment. Data from these detectors were previously used in a spin-dependent DM search in Ref.~\cite{PhysRevD.106.092008}. Two of the detectors, called Li1P and Li2P, are optimized to collect athermal phonons produced by nuclear recoils within their absorber crystals made of lithium aluminate. The third detector, Li1L, uses a silicon-on-sapphire (SOS) wafer to collect the scintillation light produced by particle recoils in the scintillating target of Li1P. Li1P and Li1L are operated within a joint housing. However, for this work, we will treat them as independent detectors. We show in Fig.~\ref{fig:pulse_noise} exemplary for Li1P the comparison between measured and simulated transition curve, pulse shape, and NPS, which all agree to a satisfying degree for our purposes. In \ref{app:parameters}, the physics parameters extracted from the measurement and used for the simulation are summarized, as well as further details about the used functions and process.

\subsection{Training}

\begin{figure*}[ht!]
\centering
\includegraphics[width=\linewidth]{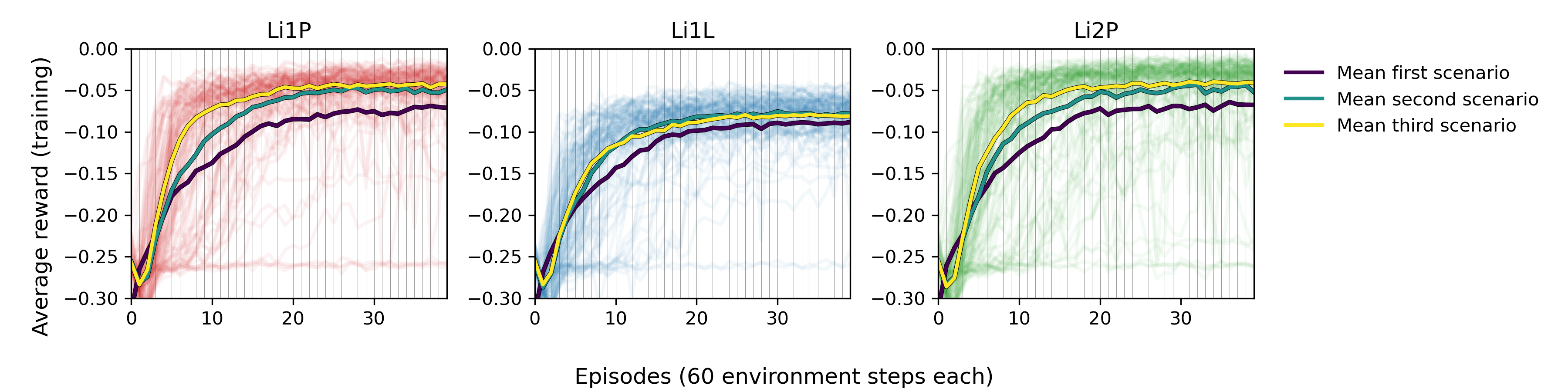}
\caption{ The average rewards per episode during training for all 105 versions of the three detectors Li1P (red, left), Li1L (blue, center), and Li2P (green, right). The thick lines are the mean values of all curves corresponding to the first/second/third scenario (violet/turquoise/yellow). The mean values rise close to the apex of the curves after 15 to 20 episodes. The second and third scenarios reach convergence significantly faster than the first. During the first 5 to 10 episodes, only little return is collected. The distribution of curves is clearly not normally distributed around the mean value, which is due to the different hyperparameter settings in the training of the individual detector versions.}
\label{fig:rewards_virtual}
\end{figure*}

We tested the optimization of control parameters with RL in a virtual environment. For this, we wrapped the simulation of the three CRESST detectors introduced in Sec.~\ref{sec:setup} in an OpenAI\footnote{The main development of this open source project was recently transferred to a fork maintained by the Farama foundation.} Gym environment \cite{brockman2016openai} and defined actions, state, and reward as described in Sec.~\ref{sec:Analysis}. We sent test pulses in increasing order, with $TPA$ values containing all integers from 1 to 10 and the values 0.1 and 0.5. After each test pulse, the agent could adjust the control settings, which jointly represents one environment step. We ran episodes of 60 environment steps and reset the detector to a randomly chosen value on the edge of the control parameter space at the start of each episode. The value ranges included in the control parameter space are chosen from reasonable experience and are not tuned to individual detectors. One environment step corresponds to the equivalent of 10 seconds of measurement time on the CRESST setup. 

For each of the three detectors, we tried 3 scenarios of the training procedure and behavior of the environment. For each scenario, we trained with 7 different hyperparameter settings and generated for each of those combinations 5 detector and agent versions. The detector and agent versions differed from each other by an individual choice of the random seed for the stochastic training procedure. Additionally, we randomized the physics parameters of the simulation by 20 \% of their original value to reduce the impact of stochastic fluctuations in the training and detector parameters. The subset of hyperparameters that were changed in the 7 different settings are the learning rate and batch size for the training of the neural networks, the discount factor $\gamma$ for the RL training, and the number of gradient steps for which the neural networks are trained on the replay buffer after each environment step. The first setting corresponds to the default setting chosen in Ref.~\cite{haarnoja2019soft} but with a lower batch size and a higher number of gradient steps. For each of the detectors, an individual SAC agent was trained for 35 versions in three different training scenarios, leading to a total of 105 versions for each detector. One such training lasts 40 episodes. The three scenarios are: 
\begin{enumerate}
    \item In the first scenario, we apply the procedure as outlined above without any adaptions. 
    \item In the second scenario, we perform a fast sweep in the control parameter space before the training starts and add the collected experience to the replay buffer. This sweep is done by gradually lowering the $DAC$ value from its maximal value to zero while the $IB$ value oscillates for each $DAC$ value, either from its highest value to zero or vice versa. In total, 120 environment steps are spent in the initial sweep. 
    \item In the third scenario, we study an effect that is expected in real-world environments: temperature changes in cryogenics generally take place slowly, as additional components of the structures surrounding the detector might be impacted by the heating on much larger time scales than the observed pulse shapes. The potential impact of this handicap is simulated by delaying the effect of the constant heating controlled by the $DAC$ value on an exponential time scale of 20 seconds, the equivalent of two environment steps. This delay is also implemented for the other scenarios, mostly to stabilize the behavior of the numerical ODE solver, but set to a value of 1 second, which has no observable impact on the time scale of environment steps.
\end{enumerate}
\noindent The resulting average reward within the training is shown in Fig.~\ref{fig:rewards_virtual} for all versions and detectors. For the large majority of trained detector versions, the return settles on a high value after 15 to 20 episodes, which indicates that an optimal OP is found. This exploitation period is preceded by an exploration period until the agent finds the superconducting transition and a good OP within it. More technical details for implementing and training the SAC agents can be found in \ref{app:training_virtual}. The different choices of hyperparameters are studied in more detail in \ref{app:training_hyperpar}. The overall conclusion is that a higher number of gradient steps, a higher learning rate, a lower $\gamma$ value, and a higher batch size seem to have a positive impact on the speed of convergence but also lead to a higher risk of failure, i.e.~that no good OP is found after training. This is expected since all these interventions shorten the exploratory period of the training. 

\begin{figure}[!b]
\centering
\includegraphics[width=\linewidth]{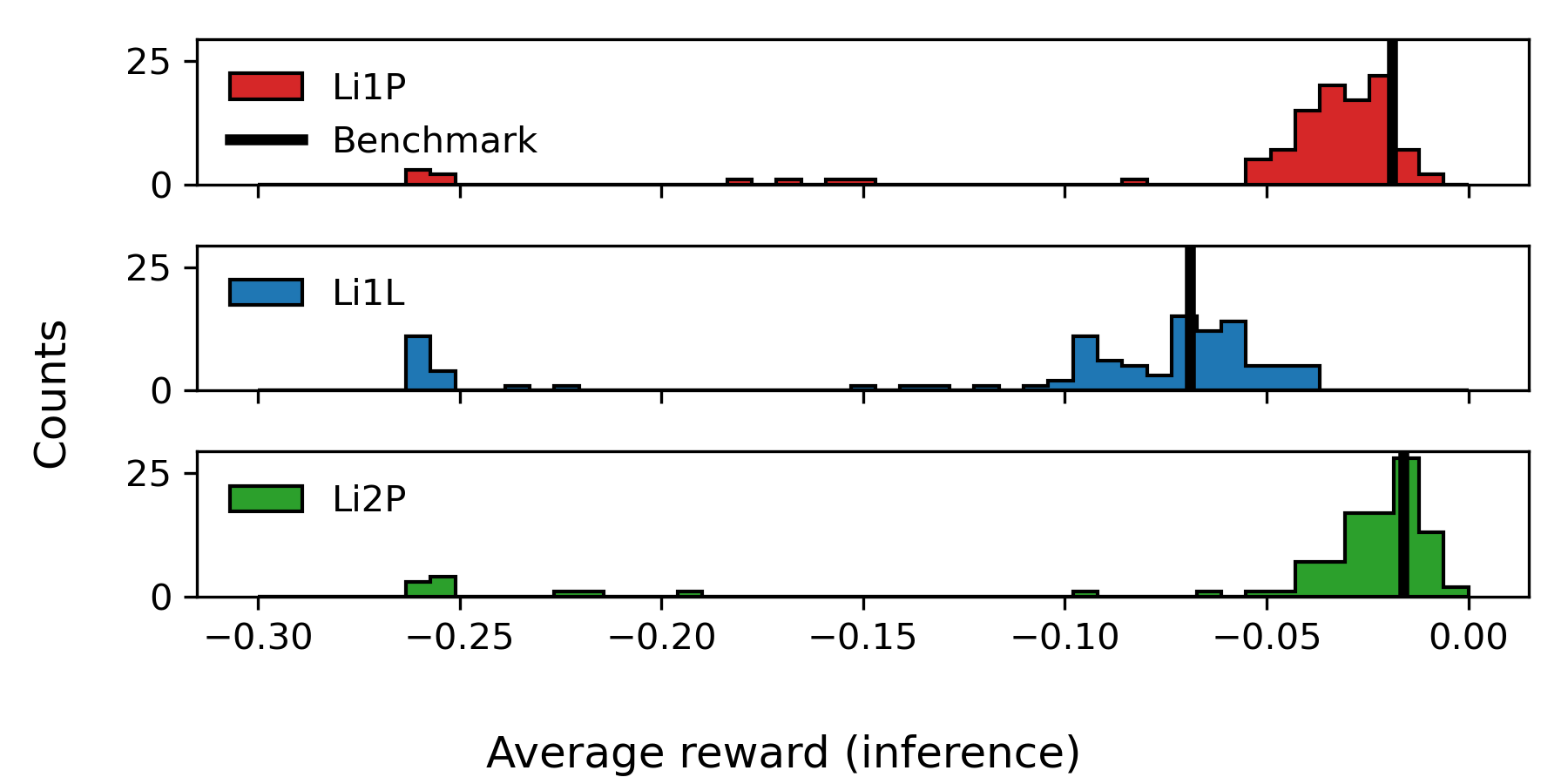}
\caption{Histogram of the average reward achieved during inference trajectories with the trained agents for the 105 versions of Li1P (red, top), Li1L (blue, center), and Li2P (green, bottom) each. Rewards from versions with opportune choices of hyperparameters cluster around a benchmark value (black line), achieved by a human expert. The suboptimal versions appear at lower reward values in the histogram. The results of Li1L surpass the benchmark value, since higher pulses saturate stronger in this detector, which can be accounted for with the machine learning method.}
\label{fig:rewards_sac_sim}
\end{figure}

\subsection{Results}

\begin{figure*}[!t]
\centering
\includegraphics[width=\textwidth]{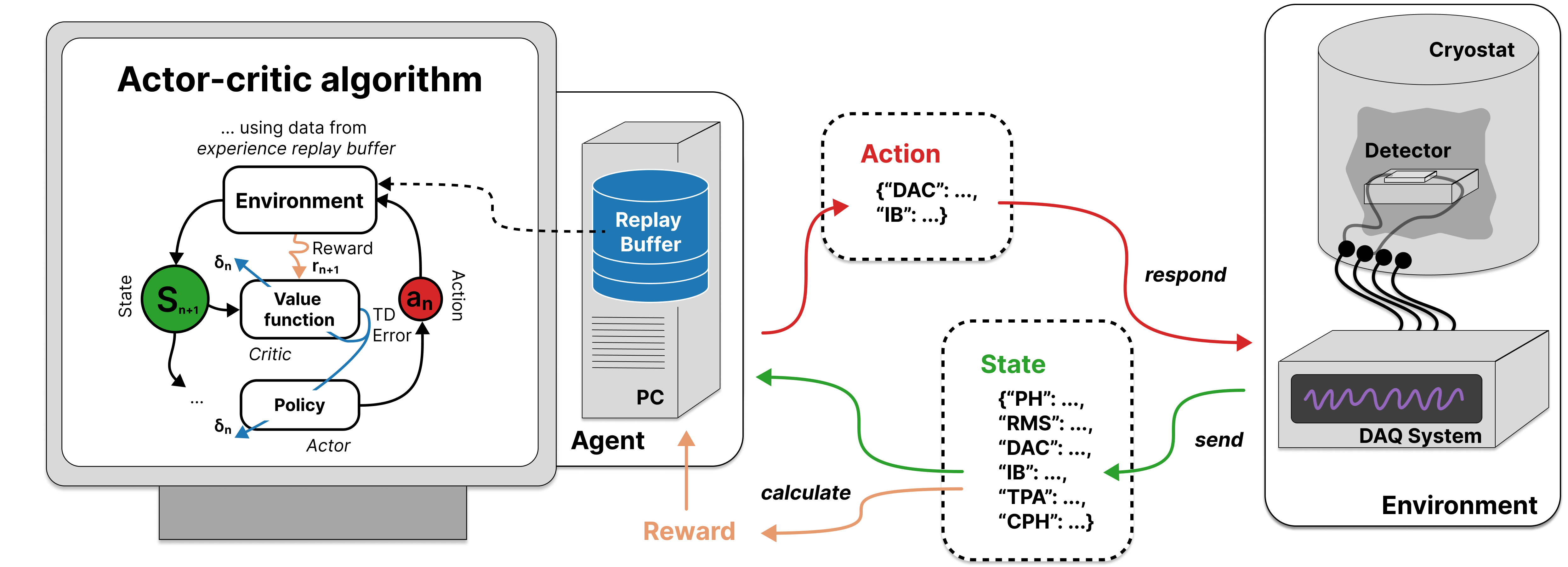}
\caption{Schematic visualization of the implemented setup to optimize CRESST detector control. (right side) The detectors are operated in a cryostat and read out by a DAQ system. The parameters of recorded test pulses are sent via an MQTT broker to a client, as state. (left side) The client calculates the reward from the state, stores the data in an experience replay buffer, and responds to the DAQ system with new control parameters. An independent process trains the AC agent on the buffer. This is a symbolic visualization, the algorithm we are using is the SAC algorithm. }
\label{fig:live_rl}
\end{figure*}

We call trajectories "inference trajectories" when we take the expectation of the Gaussian policy function instead of sampling from it to test the performance of the trained agent. In such trajectories, the agent moves directly to the learned optimal OP, which is usually close to the steepest point of the superconductor's transition curve. In our simplified models of the curves, this is at half of its normal conducting resistance. 

We have observed that the SAC agent changes the control parameters after each pulse. This behavior can result from one of two situations: first, it is possible that multiple OPs in the parameter space are equally optimal, in which case the random initialization of the neural networks determines which action the agent prefers to take. A trick to prevent this random choice of actions is to penalize the agent for switching the OP. We experimented with adding an additional regularisation term to the reward function, which penalizes large steps in the parameter space, by subtracting the Euclidean distance between current and new OP, multiplied by a hyperparameter $\omega$. The response of the agent to this regularization is delicate. It did not change the behavior visibly for small values of $\omega$, and had a negative impact on the exploratory behavior of the agent for large values of $\omega$. We found a suitable trade-off by using the value $\omega=0.1$. This behavior and the regularization strategy are discussed in more detail in \ref{app:regularization}. 

A second explanation for the changing choice of OPs is that different OPs are optimal for different magnitudes of injected test pulses. This behavior is expected since we know that our formulation of rewards is a trade-off between low noise and linear detector response, and larger injected pulses are more likely to exceed the region of linear detector response. This switching between OPs is studied in more detail with the experiments conducted on the live CRESST setup in Sec.~\ref{sec:live}, to prevent effects that only emerge in the virtual environment. We found that both effects explained above likely impact the choice of action and jointly cause the switching between OPs during inference trajectories. However, this is not a limitation of our method since one can always fix the OP corresponding to a specific magnitude of injected test signal for the final operation of the experiment, the physics data taking. On the contrary, it allows one to choose an OP that provides the desired trade-off between linearity and low noise conditions for a certain recoil energy of interest. A possibility to generally prevent this behavior would be to redefine the environment step not as the injection of one test pulse but as the injection of the whole sequence of test pulses, calculate the reward, and choose a new action only after all test pulses are injected. While this would certainly prevent the switching of the agent between OPs due to different optimality for different injected $TPA$s, acquiring a suitable amount of experience in the environment would take much longer. Another possibility is to randomize the order of the test pulses, which would leave only jumps due to the random initialization of the neural networks. 

We tested the optimality of the OPs that the SAC agents found by comparing them with a benchmark value that a human expert for the physics data taking of a previous measurement period found. We calculated the average reward of 60 test pulses from the previously recorded data with the benchmark OP. Furthermore, we ran inference trajectories with the trained SAC agents on their individual detector versions and recorded the average reward during those inference trajectories. The results are visualized in Fig.~\ref{fig:rewards_sac_sim}. The comparison between different detector versions and the benchmark value is subject to uncertainties. First, randomizing the physics parameters in the detector versions also leads to a randomization of the overall achievable energy threshold and average reward. Second, the results shown in Fig.~\ref{fig:rewards_sac_sim} include all training and environment scenarios and choices of hyperparameters, while some of them lead to systematically better or worse results than others. Third, the benchmark values were obtained by the effort of a team of several human experts and recorded during physics data taking. In this period, control settings were not adjusted over a long time period, leading to stable equilibrium conditions. On the contrary, the RL agent frequently adjusts the control settings, which can cause long-term thermal relaxation processes and, therefore, less stable noise baselines and a higher $RMS$ value. However, we can state the overall observation that the SAC agent reaches, given suitable choices of training hyperparameters, similar average rewards as the human experts could.

\vspace{1cm}

Overall, it was shown in this section that SAC agents can be trained to find optimal OPs for TES-based cryogenic calorimeters within our virtual environment. The required equivalent measurement time varies with the chosen hyperparameters and detector but can be estimated to be several hours. While an expert can likely perform this task equally fast by hand, our procedure can be parallelized for all operated detectors and executed during time periods when manual interactions are cumbersome, e.g.~at night. In the following section, we validate our method by operating on the real-world environment of the CRESST experimental setup. While all experiments in this section were conducted with standard detector designs, our method can also be applied to more complex designs with more interacting control parameters. This is shown in \ref{app:multicomponent}, where we train on virtual detectors with two TES.

\section{Live operation on the CRESST setup} \label{sec:live}

\textit{We interfaced our reinforcement learning framework with the setup control software of the CRESST experiment and performed the first proof-of-principle measurements in the real world. We tuned the algorithms for fast and reliable convergence and reached convergence within approximately 1.5 hours of measurement time in all six optimization runs, which is faster than the typical human expert. The optimality of the found operation points is well within the expectation compared to our results in the virtual environment. We discuss how the algorithms can be tuned in follow-up measurements to reach the optimality achieved by human experts.}

\begin{figure}[!t]
\centering
\includegraphics[width=\linewidth]{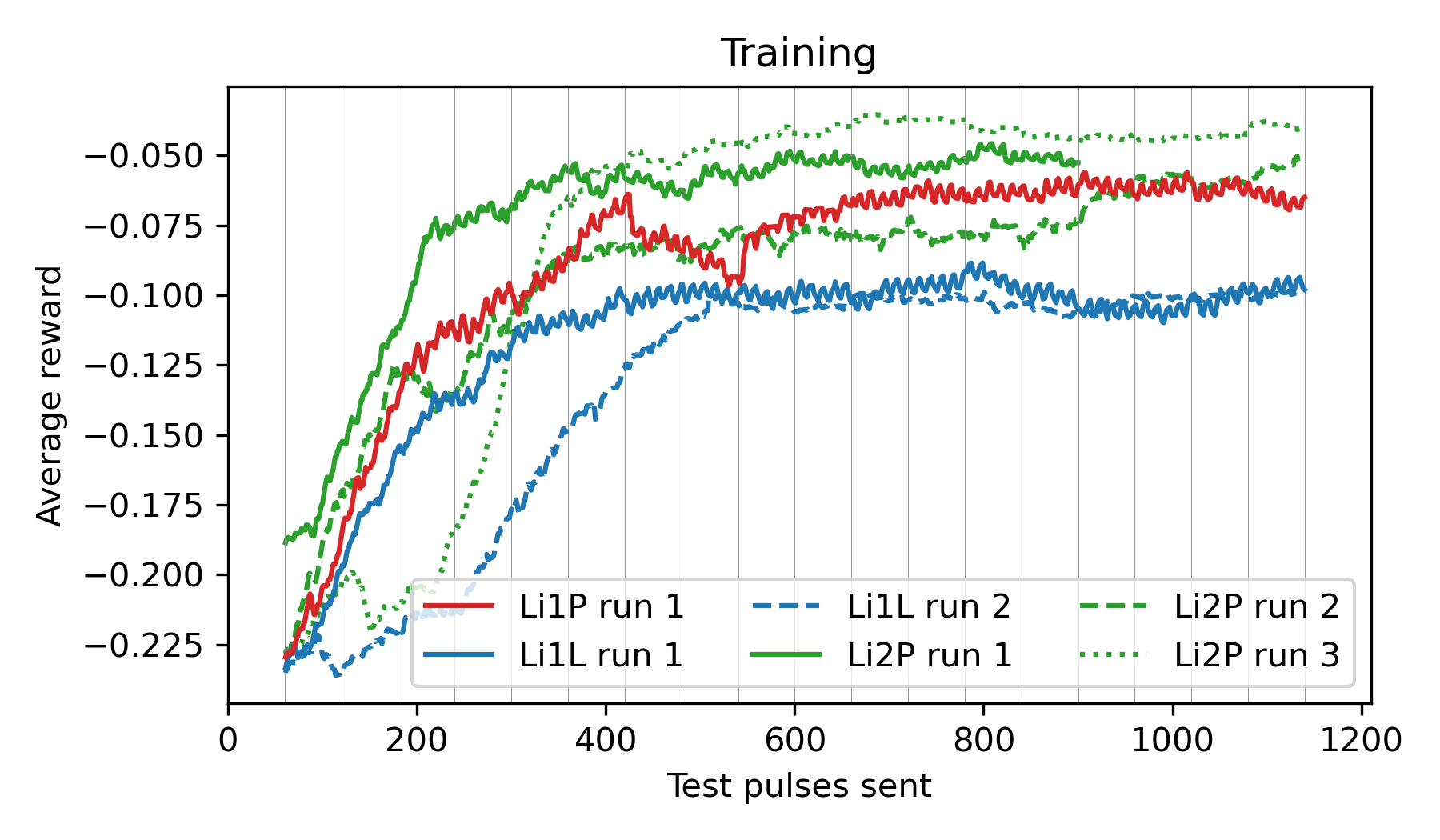}
\caption{Average rewards per test pulse sent during the live training on the CRESST setup, smoothed with a moving average of 60 test pulses. Results from six runs with different training settings are shown for Li1P (red), Li1L (blue, blue dashed), and Li2P (green, green dashed, green dotted). For this comparison, we re-calculated the rewards after training with Eq.~(\ref{eq:adapted_target}), while during training for some of the runs, the unweighted reward function Eq.~(\ref{eq:raw_target}) was used.}
\label{fig:rewards_live}
\end{figure}

\begin{figure*}[!t]
\centering
\includegraphics[width=0.7\linewidth]{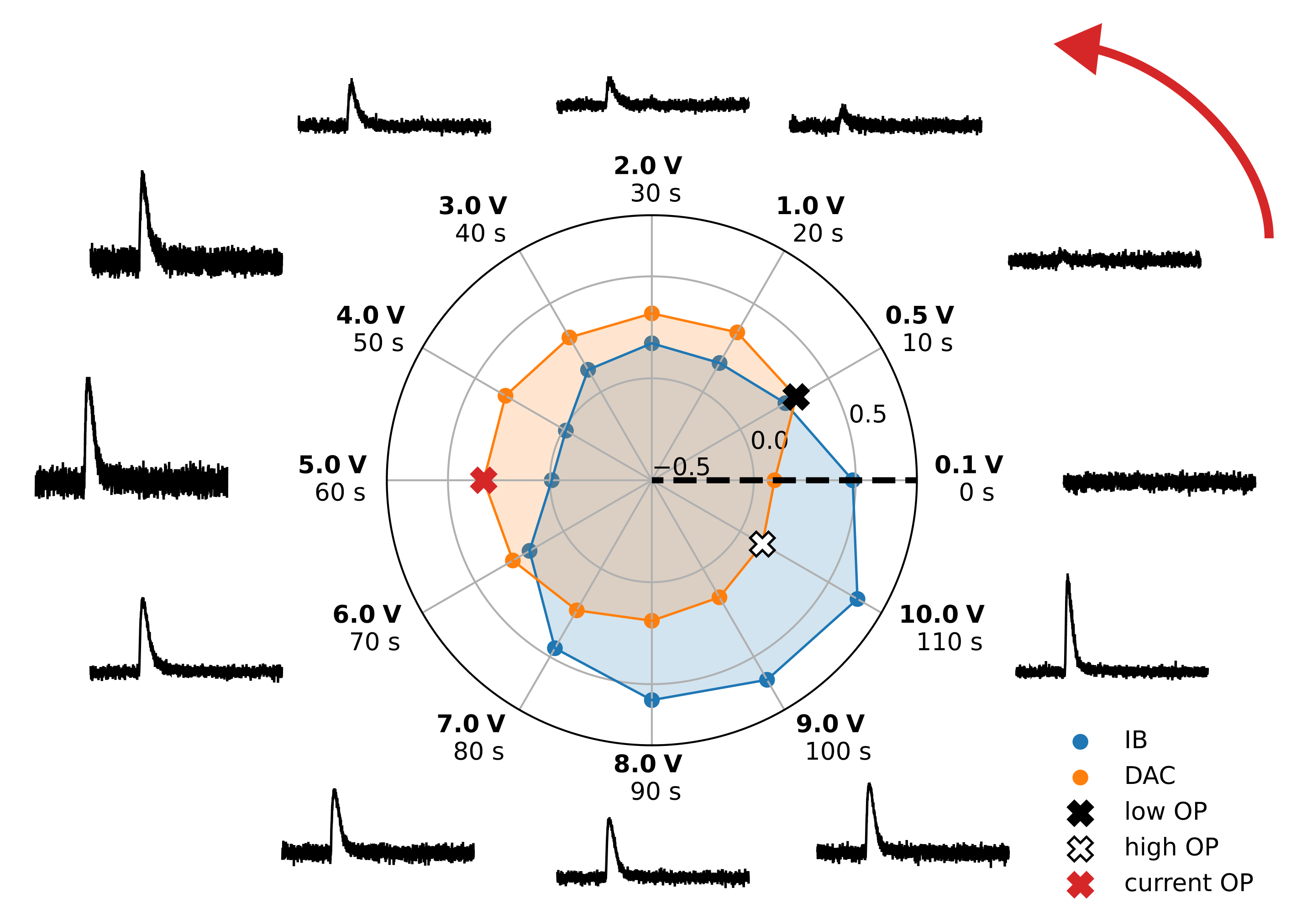}
\caption{Visualization of the cyclic adjustment of the control parameters during an inference trajectory on Li1L, run 2. The ascending trajectory of injected test pulses is visualized in the circle in the anti-clockwise direction. The voltage traces of the observed pulses (black) are normalized to a fixed voltage interval. The pulses are normalized to the applied bias current, leading to smaller pulses and noise for higher $IB$. The $TPA$ values (bold) and measurement time since the start of the test pulse trajectory are written next to the voltage traces. The polar plot includes the $IB$ and $DAC$ values that were set while the corresponding pulse was recorded. Their values are normalized to the interval -1 to 1 (see Sec.~\ref{app:training_live} for normalization ranges). The polar axis starts at -0.5, and the distance between the grey rings corresponds to an increase of 0.5. Three OPs are marked with black, red, and white crosses, corresponding to OPs that were chosen for low, intermediate, and high $TPA$ values. }
\label{fig:circle_live}
\end{figure*}


A measurement interval of 12 days in February 2023 was dedicated to testing the method of optimizing detector operation live on the CRESST underground setup in the Laboratori Nationali del Gran Sasso (LNGS). Experiments were performed with Li1P, Li1L, and Li2P, the three detectors of which virtual twins were described in Sec.~\ref{sec:setup} and used for RL experiments in virtual environments in Sec.~\ref{sec:virtual}. 

The communication between our Python-based RL infrastructure and the control and data acquisition (DAQ) software of the CRESST setup was realized via messages sent through an MQTT broker, which is a widely used internet-of-things communication protocol. The DAQ system, which acted as the RL environment, induced test and control pulses through the heater electronics and recorded the detector response. Pulse shape parameters were calculated, and a message was broadcast via the broker and received by the machine on which we ran the RL infrastructure. On this machine, we ran two parallel processes: 
\begin{enumerate}
    \item The first process received messages, calculated rewards, and wrote data to the replay buffer. A policy model was queried with the state of each received message. The outputs were compiled into a reply containing new control settings.
    \item The second process continuously trained the agent with a SAC algorithm on the experience replay buffer. The process was paused if the desired number of gradient steps was reached before new data was added to the replay buffer. The accessibility of the replay buffer from both processes was realized through memory-mapped arrays. 
\end{enumerate}
This setup is schematically visualized in Fig.~\ref{fig:live_rl}. We ran experiments consecutively as our current implementation of device communication does not support work on multiple channels in parallel. A total of 48 experiments were run with measurement times between one and three hours, where the majority was used for implementation and debugging of the setup, and the final 6 runs were used as performance benchmarks of the method. 

One performance run was performed with Li1P, two with Li1L, and three with Li2P. Each run was started with a rough sweep of the action space, as done in scenarios 2 and 3 of the detector versions in Sec.~\ref{sec:virtual}. Furthermore, we observed delays until strong changes in the heating took place, as modeled in scenario 3 of the virtual versions. We made individual adaptions to the hyperparameters, the configuration of the state space, and the number and length of training episodes in all runs. These and further details of the training process are summarized in \ref{app:training_live}. The objective of the choice of hyperparameters was to obtain a proof-of-principle by achieving fast and reliable convergence of the algorithms, with less emphasis on the optimality of the found OPs. This is mainly represented by the smaller choice of the regularization parameters $\omega$, and the constant (instead of decreasing) target entropy. 
The average rewards obtained during training, depending on the number of test pulses sent since the start, is shown in Fig.~\ref{fig:rewards_live}. In all runs, a high plateau of rewards is reached before 600 test pulses were sent, corresponding to roughly 1.5 hours of measurement time. In comparing the runs, we have qualitatively observed that a state space containing more variables leads to a longer required time until an optimal OP is found, but also to generally better responsitivity to the environment. 
After training is completed, we run inference trajectories with all trained agents, i.e.~we choose the expectation of the Gaussian policy function instead of sampling from it. They all find suitable OPs and feature a similar behavior that is exemplary visualized for Li1L run2 in Fig.~\ref{fig:circle_live}. We observe that the agent adjusts the OP to the $TPA$ value of the injected test pulse expected next in the cyclic test pulse queue. The same observation was made in our virtual environment in Sec.~\ref{sec:virtual}. By comparing the recorded noise traces, we can see that the agent prefers an operation point with high $IB$ and low $DAC$ for small injected pulses but one with a higher $DAC$ and lower $IB$ for larger injected pulses. We interpret this behavior as the existence of two different optimal trade-offs between the linearity of the detector response and noise conditions, between which the agent cyclically switches, depending on the $TPA$ of the pulse. If an approximate energy calibration of the detector can be performed, i.e.~a relation between the recoil energy $\Delta E$ and the $TPA$, the policy can be used to optimize the control parameters of the detector for the $TPA$ and $\Delta E$-region of interest. As a potential limitation, the precision of this energy calibration can itself depend on the OP.

\begin{figure}[!b]
\centering
\includegraphics[width=\linewidth]{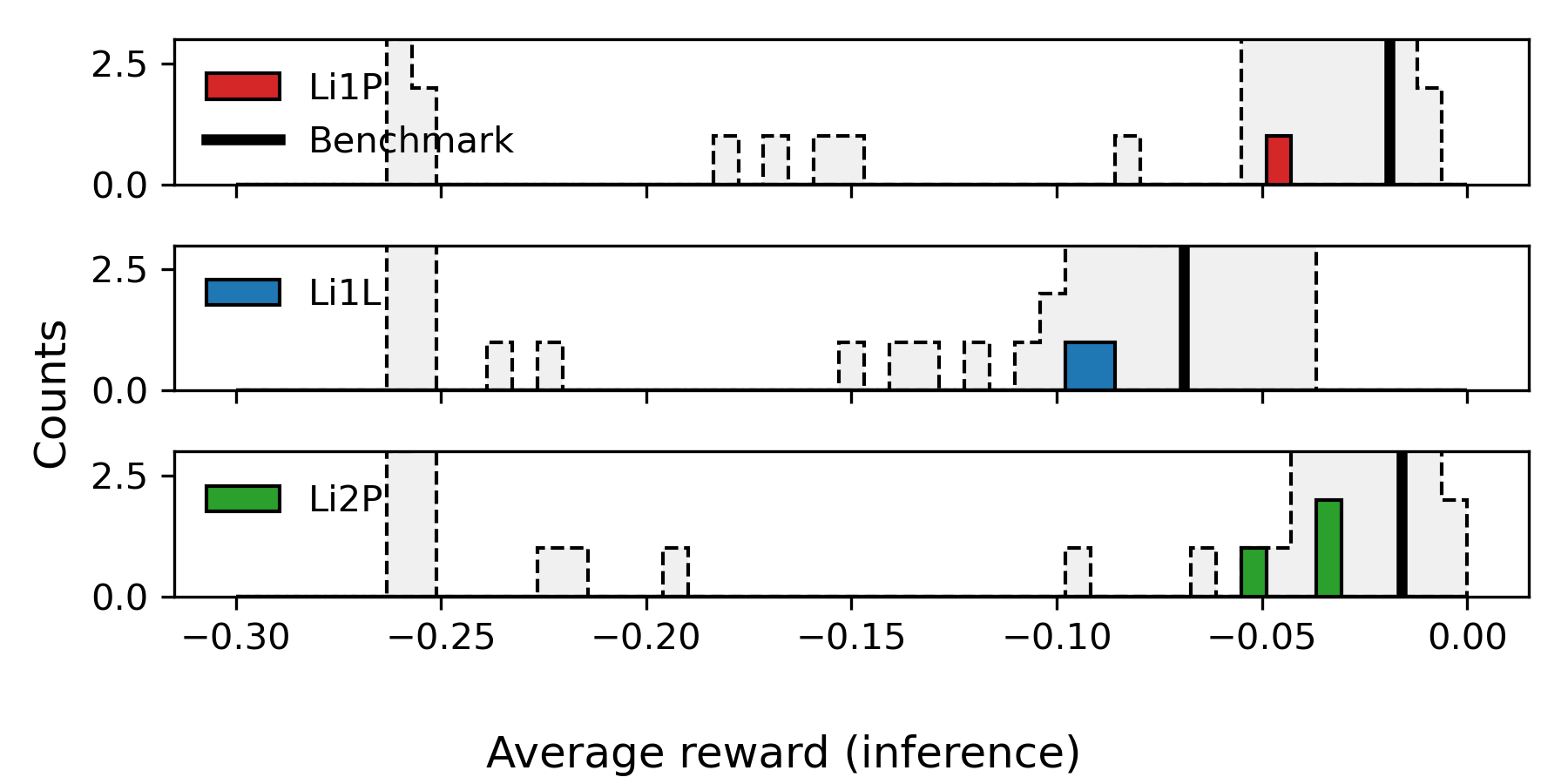}
\caption{Histogram of the average reward obtained during inference trajectories with the trained agents on the real-world versions of Li1P (red, top), Li1L (blue, center), and Li2P (green, bottom) each. The rewards obtained in the simulation (grey, dotted histogram) and the human-optimized benchmark value (black line) are shown for comparison. The obtained rewards are worse than the benchmark value but correspond to our expectations from the simulation. For a discussion of the achievable optimality, see also Fig.~\ref{fig:live_values}.}
\label{fig:rewards_sac_real}
\end{figure}

The optimality of the OPs found with RL, quantified as the average reward during inference trajectories, is shown in Fig.\ref{fig:rewards_sac_real}. The rewards obtained in the virtual environment and the human-optimized benchmark value are compared. The comparison is subject to the same uncertainties as stated in Sec.~\ref{sec:virtual}. Additionally, it is possible that the overall noise conditions of the detectors changed since the data-taking period during which the benchmark values were obtained since several warm-up tests were performed on the CRESST setup in the meantime (see Ref.~\cite{10.21468/SciPostPhysProc.12.013}). Since the algorithms were not tuned to achieve the highest possible rewards but fast convergence, it does not come as a surprise that the optimality is behind the human-optimized benchmark values. However, the obtained rewards are within the distribution of rewards obtained in the simulation. Therefore, we expect that tuning the algorithms towards optimal rewards will lead to similar results as we observed in the virtual environment by reaching the human-optimized benchmark values. The technical reason for our expectation is discussed in the remainder of this section.


\begin{figure}[!t]
\centering
\includegraphics[width=\linewidth]{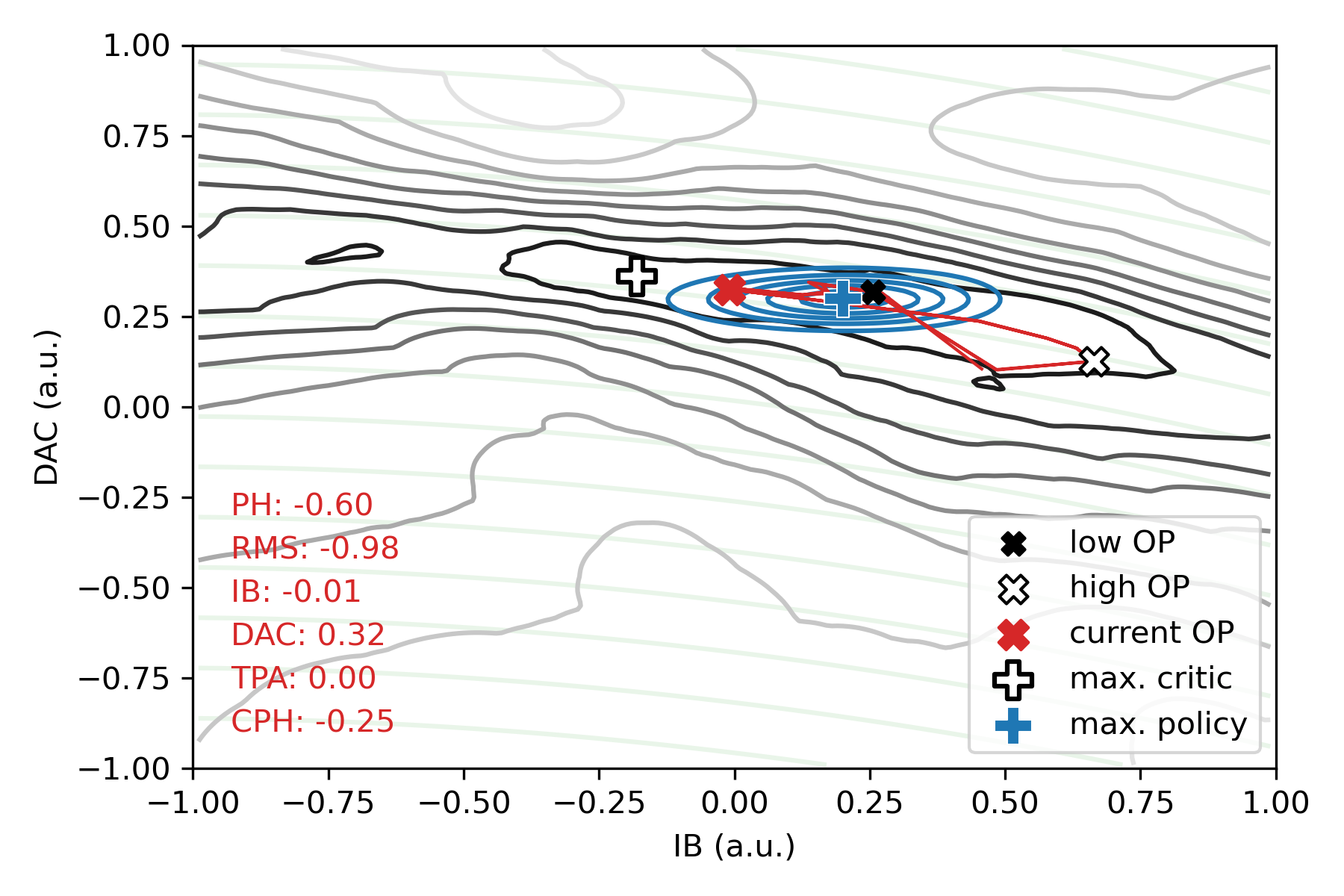}
\caption{Visualization of the Gaussian policy probability distribution (blue) and the critic function (grey-black) over the two-dimensional action space, for a fixed "current" state (red text, lower left) and Li1L run 2. The maximum of the critic function is marked with a white plus. The current control parameters are marked with a red cross, that of OPs that were chosen by the agent for high/low $TPA$ values with a white/black cross. These crosses correspond to the OPs marked with similar crosses in Fig.~\ref{fig:circle_live}. The trajectory of actions that are chosen by the agent in inference is drawn with a red line, partially covered by the blue policy function. We can clearly see a mismatch between the actions preferred by the policy function and the maximum of the critic function. The reason for this mismatch is discussed in the text and in \ref{app:entropy}. The expected lines of constant heating caused by the $DAC$ through the heating resistor and the $IB$ through Joule heating are shown in the background (light, transparent green). As expected, the island of actions that the critic prefers stretches along the constant heating lines and corresponds to a fixed resistance of the TES. The state values are normalized to the interval -1 to 1. The original value ranges are written in Tab.~\ref{tab:live_hyperpars}.}
\label{fig:live_values}
\end{figure}

In Fig.~\ref{fig:live_values} we study the critic and policy function of the SAC that was trained on Li1L, run 2. We observe a severe mismatch between the chosen actions and the maximum of the critic function. This behavior is counterintuitive since we would expect that the policy has learned to choose the actions that are identified as most opportune by the critic once both neural networks have converged. However, the SAC algorithm is intrinsically built on the stochasticity of its policy function. The policy does not learn to position its expected value on top of the maximum of the critic, but it converges to a region of the action space where randomly sampled actions from its probability distribution are expected to have maximal values. Since we simultaneously force the Gaussian policy function to have a certain target entropy, it cannot converge to features of the critic function on smaller value scales than its entropy allows. Therefore, it is impossible for the policy function to converge towards the maximum of the critic function and the best possible operation point. This behavior is further discussed on a simpler toy model in \ref{app:entropy}. In our experiments in the virtual environment, we avoided this behavior by lowering the target entropy during the training, allowing the Gaussian policy to shrink to smaller value ranges. It was an intentional choice not to apply this entropy scheduling in our experiments on the real-world setup since it could potentially have increased the risk of failure, effectively counteracting the goal of the performance runs, namely a proof-of-principle of convergence in the real world. We expect that implementing a similar entropy schedule on the live setup will lead to similar optimality as was obtained in the virtual environment and by the human expert. 

\vspace{1 cm}

In summary, we have provided proof-of-principle that RL is a practically applicable method for finding optimal control parameters for cryogenic TES-based calorimeters in the real world. Our studies from the virtual environment discussed in Sec.~\ref{sec:virtual} generalized well to the results obtained on the CRESST setup.
The time period needed for training is comparable to or shorter than the time period that a human expert needs to do the optimization via manual interventions. 
The achieved average rewards in inference trajectories are within the distribution of average rewards obtained in the virtual environment but slightly worse than the benchmark values. It is expected that similar optimality as in the virtual environment can be achieved by implementing a suitable entropy schedule and choosing suitable hyperparameters of the algorithm.

\section{Conclusion}
\label{sec:Conclusion}

In this work, we present studies for automating the optimization of control parameters of TES-based cryogenic calorimeters with RL. We simulated the response and noise of three CRESST detector modules and trained SAC agents to find optimal OPs for them. We sampled randomized versions of the detectors and systematically studied hyperparameters of the training and RL setting on a total of 105 different versions of each detector. We tested our method on the real-world versions of the detectors operated in the underground CRESST setup in six representative test runs across the three detectors. In all our experiments, the required equivalent measurement time to complete training was between 1 and 3 hours, fast enough for practical usage. On the live setup, convergence was reached in all runs within 1.5 hours, which is faster than most human experts can perform the task. The training was successful on all representative runs on the live setup and most runs in the virtual environment. The found OPs reach the optimality of a human-optimized benchmark value in the virtual environment. The tests on the live setup were not dedicated to achieving the highest possible rewards but emphasized stable convergence. We discussed the necessary adjustments to the algorithm to achieve similarly optimal OPs on the live setup as human experts achieved.

For future work, the presented method could be combined with more control parameters, e.g.~that of an active magnetic field compensation, since the magnetic field can also influence the shape of the transition curve. Furthermore, richer information extracted from the observed pulse shape could improve the agents' stability and convergence speed, e.g.~a combination with networks that discriminate pulses from artifacts and pile-up as reported in Ref.~\cite{angloher_towards_2023}. In stable measurement setups, the rate of test pulses can be increased, reducing the required measurement time. It is also conceivable to replace the injected test pulses with a sinusoidal signal in the relevant frequency range of pulses to directly measure the SNR by separating the injected frequency from other noise with frequency filters. On the algorithmic side, we could combine more of our prior knowledge about TES detectors with the model-free SAC algorithm. Currently, our method's main risk of failure is that the agent might not find the island of suitable $DAC$ and $IB$ value combinations, which correspond to the superconducting transition, in its exploration period. This risk could be lowered by forcing the agent to sample from unexplored regions of the action space until a certain reward is obtained (e.g. a value above -0.15) or by extending the period of the initial sweep. Furthermore, we did not explicitly use the knowledge that a change in the noise conditions should be visible whenever the superconductor switches from a normal to a superconducting state. 

In summary, the presented method can significantly reduce the required time for the initial control parameter optimization of large multi-detector setups, thereby increasing the time available for physics data-taking. Together with the deep learning method published previously by the CRESST collaboration \cite{angloher_towards_2023}, for automated data cleaning for cryogenic detectors, we can dramatically reduce the human workload required for tuning of a scaled-up number of cryodetectors. Furthermore, we expect some of the learnings we describe to be transferable to the general problem of automating experiments with RL. We have demonstrated that cryogenic detectors, with their unique characteristics such as hysteresis and noise peculiarities, can be efficiently tuned on a large scale using RL, achieving the precision and speed of human experts. This promising application of RL opens the door to its extensive exploration in other fundamental physics experiments that operate with a high number of channels.

\begin{acknowledgements}
We are grateful to LNGS-INFN for their generous support of CRESST. This work has been funded by the Deutsche Forschungsgemeinschaft (DFG, German Research Foundation) under Germany's Excellence Strategy – EXC 2094 – 390783311 and through the Sonderforschungsbereich (Collaborative Research Center) SFB1258 ‘Neutrinos and Dark Matter in Astro- and Particle Physics’, by the BMBF 05A20WO1 and 05A20VTA and by the Austrian Science Fund (FWF): I5420-N, W1252-N27 and FG1 and by the Austrian research promotion agency (FFG), project ML4CPD. The Bratislava group acknowledges a support provided by the Slovak Research and Development Agency (projects APVV-15-0576 and APVV-21-0377). The computational results presented were partially obtained using the Vienna CLIP cluster and partially using the LNGS computing infrastructure. Pulse shape processing was performed using the Cait open source package \cite{wagner_cait_2022}.
\end{acknowledgements}

\appendix

\section{Glossary} \label{app:glossary}

This glossary contains only the most important terms from the paper that may not be familiar to all readers.

\begin{itemize}
    \item \textbf{Action:} A combination of control parameters that the agent chooses to set. 
    \item \textbf{Agent:} The RL model that interacts with the environment through actions, states, and rewards. Contains, in our case, a policy and a value function.
    \item \textbf{Bias current}, $IB$: The current applied to the readout circuit of the TES. One of two control parameters in the RL optimization method.
    \item \textbf{Environment:} A system that interacts with the agent by receiving actions and returning states and rewards. In the context of Sec.~\ref{sec:virtual}, this is a simulation of a detector. In Sec.~\ref{sec:live}, this is the physical detector and DAQ system. 
    \item \textbf{Episode:} The training of an RL agent is split into finite episodes. In our work, an episode contained 60 time steps.
    \item \textbf{Heater:} A resistor placed on the target crystal of the detector to control the temperature and inject test pulses. 
    \item \textbf{Heating}, $DAC$: A quantity that controls the constant component of the current applied to the heater. Its main purpose is to control the temperature of the TES.
    \item \textbf{Noise width}, $RMS$: The root-mean-square value of the pre-trigger region of a record window. This value is a good estimate of the current noise conditions of the detector. 
    \item \textbf{Operating point, OP}: The combination of temperature, resistance and bias current in which the TES resides when no pulses are observed. We use this synonymously to refer to a combination of control parameters $DAC$ and $IB$.
    \item \textbf{Policy/Actor:} A function that maps a state to an action. Learning an optimal policy from the observed data, one that maximizes the return, is the agent's goal.
    \item \textbf{Pulse height}, $PH$: The maximum of a record window with the mean value of the pre-trigger region subtracted.
    \item \textbf{Regularizer:} An additional, additive term in the reward function that should prevent or emphasize certain behaviors of the trained model. Usually, the value of the regularizing term is intentionally kept small compared to the main target function. 
    \item \textbf{Replay buffer:} A storage of previously observed action-state transition.
    \item \textbf{Reward:} A scalar value obtained by the agent from the environment in response to an action.  
    \item \textbf{Return:} The sum of obtained rewards. 
    \item \textbf{Soft Actor-Critic, SAC:} A specific algorithm that determines the design and details of the training for an agent.
    \item \textbf{State:} A vector-valued quantity obtained by the agent from the environment in response to an action. 
    \item \textbf{Test pulse amplitude}, $TPA$: The amplitude of the injected test pulses. 
    \item \textbf{Transition curve:} The complex dependence of the TES resistance $R_f$ on the temperature. This curve is almost flat above the transition temperature $T_c$, and zero below it. It has a non-trivial shape close to the transition temperature. 
    \item \textbf{Transition-edge sensor, TES:} A superconducting film operated in the transition from superconducting to normal conducting state and used as a highly-sensitive temperature sensor.  
    \item \textbf{Value function/Critic:} A function that maps action-state pairs to estimates of the future return.
\end{itemize}

In our notation, we keep all values that are contained in the RL action- or state space capitalized, without subscript and italic, e.g.~$IB$ for the bias current. Variables that appear in formulas are italic but not necessarily capitalized and may have subscripts, e.g.~$I_H$ for the heater current. 

\section{Derivation of the reward function} \label{app:reward}

The reward function stated in Sec.~\ref{sec:Analysis} can be derived as follows. Our overall objective is to minimize the energy threshold $E_{th}$. The optimization problem can be written as:

\begin{equation}
    \underset{DAC,\ IB}{\arg\min}( E_{th}).
\end{equation}

\noindent We need to estimate the energy threshold with easily accessible observables. To do so, we can exploit the fact that the noise amplitudes are small compared to the dynamic range of the detector and therefore fall in an approximately linear detector response regime. Therefore we can define a voltage-energy calibration constant $\gamma$ (small signal approximation), and replace the energy threshold with a voltage threshold $U_{th}$: 

\begin{equation}
    \dots = \underset{DAC,\ IB}{\arg\min}( \gamma U_{th}).
\end{equation}

\noindent Furthermore, the energy threshold of a physics search is usually defined as a multiple of the noise resolution, which can for our purposes be reasonably well approximated by measuring the observed noise's root-mean-square (RMS) value: 

\begin{equation}
    \dots \propto \underset{DAC,\ IB}{\arg\min}( \gamma RMS).
\end{equation}

\noindent Finally, $\gamma$ is proportional to the ratio between a set of N injected test pulses with amplitudes $TPA_i$, the observed pulse height $PH_i$:

\begin{equation}
    \dots \propto \underset{DAC,\ IB}{\arg\min}\left( \sum_{i=1}^N \frac{TPA_i}{PH_i} RMS_i\right).  
\end{equation}

\noindent The quantity $RMS_i$ refers to the noise RMS evaluated on the records of the individual test pulses. We omitted writing the functional dependencies in this derivation for better readability. All appearing quantities depend on $DAC$ and $IB$, except $TPA$. The quantity $U_{th}$ is the observed voltage pulse height corresponding to the energy threshold. The assumption of a linear response is not necessarily satisfied in practice, as the transition curve of the superconductor can saturate (flatten) when larger signals are injected. In our objective, this flattening of pulses would be equally penalized as worse noise conditions. By the choice of injected TPAs, and their frequencies, we therefore implicitly introduce a trade-off between optimal noise conditions and a linear detector response in the range of applied test pulse amplitudes. For the physics goals of light DM searches we are primarily interested in good performance for small signals, but for practical reasons, e.g.~observability of calibration lines, we want to monitor the detector response also for higher energies. Instead of injecting large test pulses less frequently we therefore choose to suppress their relevance for the optimization objective, by introducing a weight factor which we choose as the inverse signal strength $w = 1/TPA$. For usage as a target function in the following sections, we rephrase Eq.~(\ref{eq:raw_target}) to a maximization objective.

\begin{align}
\underset{DAC,\ IB}{\arg\min}\left( \sum_{i=1}^N w_i\frac{TPA_i}{PH_i} RMS_i\right) & \hspace{0.5cm} \Bigl|\text{choose } w_i=\frac{1}{TPA_i} \nonumber\\
=\underset{DAC,\ IB}{\arg\min}\left(\sum_{i=1}^N\frac{1}{\cancel{TPA_i}}\frac{\cancel{TPA_i}}{PH_i} RMS_i\right) & \hspace{0.5cm} \Bigl|\text{phrase as maximization} \nonumber\\
\equiv \underset{DAC,\ IB}{\arg\max}\left(-\sum_{i=1}^N \frac{RMS_i}{PH_i}\right) & 
\end{align}

\section{Noise contributions} \label{app:noise}

The derivation of noise contributions presented here follows Ref.~\cite{irwin_transition-edge_2005}, the model of flicker noise Ref.~\cite{galeazzi_microcalorimeter_2003}. We consider only the thermal equation of the TES and the electrical equation of its readout circuit, the first and third line in Eq.~(\ref{eq:odes}), and ignore fluctuations that might arise from the interaction of absorber and TES. These equations can be linearized in a small signal approximation. Then, equations for the resulting temperature fluctuations in the TES $\Delta T_e$ and current fluctuation in the TES branch $\Delta I_f$ for small, linear inhomogeneities, i.e.~power fluctuations $\Delta P_e$ in the TES and voltage fluctuations $\Delta U$ in the readout circuit, can be derived. These can be summarized in a transition matrix

\begin{align*}
\begin{pmatrix}
\Delta T_{e}\\
\Delta I_{f}
\end{pmatrix} =&  \begin{array}{l}
\begin{pmatrix}
s_{11}( w) & s_{12}( w)\\
s_{21}( w) & s_{22}( w)
\end{pmatrix}^{int/ext}\begin{pmatrix}
\Delta P_{e}\\
\Delta U
\end{pmatrix},\\
\end{array}
\end{align*}

\noindent where the matrix elements read differently for noise that has its origin in the TES, labeled $int$, and in other parts of the readout loop, dubbed $ext$. The relevant matrix elements are

\begin{align*}
s_{21}^{int}( w) & =-\frac{1}{I_{f0}}\Biggl[\frac{L}{\tau _{\mathrm{el}}\mathcal{L}_{\mathrm{I}}} +\left( R_{f0}-R_{s}\right) \\
&+2\pi \mathrm{i} w\frac{\mathrm{L} \tau }{\mathcal{L}_{\mathrm{I}}}\left(\frac{1}{\tau } +\frac{1}{\tau _{\mathrm{el}}}\right) -\frac{4\pi ^{2} w^{2} \tau  \mathrm{L}}{\mathcal{L}_{\mathrm{I}}}\Biggr]^{-1}, \\
s_{22}^{int}( w) & =-s_{21}^{int}( w) I_{f0}\frac{1}{\mathcal{L}_{\mathrm{I}}}( 1+2\pi wi\tau ), \\
s_{21}^{ext}( w) & =-\frac{1}{I_{f0}}\Biggl[\frac{L\tau }{\tau _{\mathrm{el}} \tau _{I} \mathcal{L}_{\mathrm{I}}} +2 R_{f0}\\
&+\mathrm{2\pi i} w\frac{\mathrm{L} \tau }{\mathcal{L}_{\mathrm{I}}}\left(\frac{1}{\tau _{I}} +\frac{1}{\tau _{\mathrm{el}}}\right) -\frac{4\pi ^{2} w^{2} \tau \mathrm{L}}{\mathcal{L}_{\mathrm{I}}}\Biggr]^{-1}, \\
s_{22}^{ext}( w) & =s_{21}^{ext}( w) I_{f0}\frac{\mathcal{L}_{\mathrm{I}} -1}{\mathcal{L}_{\mathrm{I}}}( 1+2\pi wi\tau _{I}),
\end{align*}

\noindent where we used the same definitions as in Ref.~\cite{irwin_transition-edge_2005}:

\begin{align*}
\tau  & \coloneqq \frac{C_{e}}{G_{eb}}, & \tau _{el} & \coloneqq \frac{L}{R_{f0} +R_{S}},\\
\mathcal{L}_{I} & \coloneqq \frac{I_{f0}^{2}}{G_{eb}}\left. \frac{dR_{f}}{dT_{e}}\right| _{T_{e0}}, & \tau _{I} & \coloneqq \frac{\tau }{( 1-\mathcal{L}_{I})}.
\end{align*}

\noindent The thermal or phonon noise can then be calculated by

\begin{align*}
| \Delta I_{f}(w)|_{ph}^{2} & =\left| s_{21}^{int}( w)\right| ^{2}| \Delta P_{e}|_{ph}^{2}, \\
| \Delta P_{e}| {_{ph}^{2}} &= 4k_{B} T_{e}^{2} G_{eb}\frac{2}{5}\frac{1-\left(\frac{T_{b}}{T_{e0}}\right)^{5}}{1-\left(\frac{T_{b}}{T_{e0}}\right)^{2}},
\end{align*}
\noindent the electrical Johnson noise in the TES by
\begin{align*}
| \Delta I_{f}(w)| _{Jf}^{2} & =\left| s_{22}^{int}( w)\right| ^{2}| \Delta U| _{Jf}^{2}, \\
| \Delta U| _{Jf}^{2} &=4k_{B} T_{e} R_{f0},
\end{align*}

\noindent the electrical Johnson noise in the shunt resistor, scaled by a factor $E_J$ to account for excess electrical noise, by

\begin{align*}
| \Delta I_{f} (w)| _{Js}^{2} & =\left| s_{22}^{ext}( w)\right| ^{2}| \Delta U| _{Js}^{2}, \\
| \Delta U| ^{2}{}_{Js} &=4k_{B} T_{s} R_{s} E_J,
\end{align*}

\noindent and the $1/f$ or flicker noise by

\begin{align*}
| \Delta I_{f}(w)| _{flicker}^{2} & =\left| s_{21}^{int}( w)\right| ^{2}| \Delta P_{e}| _{flicker}^{2}, \\
| \Delta P_{e} (w)| _{flicker}^{2} &=\frac{\left(\frac{\Delta R_{f,flicker}}{R_{f0}}\right)^{2} R_{f0}^{2}}{w^{\alpha }} I_{f0}^{2},
\end{align*}

\noindent where $\left(\frac{\Delta R_{f,flicker}}{R_{f0}}\right)$ and $\alpha$ are parameters to be extracted from the data. 

Additionally, the SQUID produces an amount $i_{sq}$ of white noise that is directly added to the SQUID output current, decoupled from feedback in the readout circuit. There are also peaks in the spectrum from electromagnetic interference added with height $p_0$/$p_1$/$p_2$ at the frequencies 50/150/250 Hz and scaled to the observed heights in the data as well.

The noise contributions are assumed to be Gaussian and mutually independent and can therefore be summed quadratically.

\section{Parameters used in the simulation} \label{app:parameters}

We report in Tab.~\ref{tab:pars_used} the parameters used for the simulation that was introduced in Sec.~\ref{sec:setup}.

We calculated the heat capacities for the absorbers following the Debye model, with the Debye temperature of lithium aluminate calculated from the elasticity constants from Ref.~\cite{jain_commentary_2013}, with the value 429 K, and for sapphire taken from Ref.~\cite{pantic2008}, with the value 1041 K. We evaluated them at the transition temperature $T_c$ and scaled them to the absorber volume $V_a$. 

We calculated the heat capacities for the tungsten TES with the Sommerfeld constants taken from Ref.~\cite{Kittel2004}. The values are evaluated at $T_c$ and scaled to the TES volume $V_{f}$. The stated value does not contain the increase by a factor 2.43 appearing in the transition curve, which is dynamically calculated when the differential equations are numerically solved.

A part of the tungsten film is covered with aluminum. This bilayer has a higher transition temperature compared to the uncovered tungsten, due to the proximity effect, and functions solely as an athermal phonon collector. The calculated heat capacity only includes the part of the tungsten film that is not covered by aluminum since the heat capacity of the superconducting bilayer is exponentially suppressed. 

The normal conducting resistance of the TES $R_{f0}$ and its transition temperature $T_c$ are measured values.

$R_s$, $\eta$, L, $i_{sq}$, $I_H$, $\beta$, $IB_{min}$ and $IB_{max}$ are known values of the setup. $\eta$ is the conversion factor that translates a current in the SQUID branch of the readout circuit to the observable voltage, determined by the SQUID gain settings.

The thermal couplings, as well as $\tau_n$ and $\epsilon$ were fitted to match the pulse shape of observed absorber recoils, using Eqs.~(10) and (11) from Ref.~\cite{probst_model_1995}. 

$\delta$, $\delta_H$, $R_H$, $\frac{\Delta R_{f,flicker}}{R_{f0}}$, $k$, $E_J$, $p_0$, $p_1$, $p_2$ and $\alpha$ are adjusted to match the measured data. The adjustment was done by hand since we could not obtain a satisfying fit of the system of differential equations to data with automated fitting algorithms. The parameter $k$ controls the steepness of the TES transition curve, which we simplify as a logistics function. The derivative of the transition curve at its steepest point has the value $4kR_{f0}$. The parameter $\beta$ is a scale factor between the injected test pulses and the constant heating current. The values $p_0$, $p_1$, and $p_2$ are coefficients of numerical templates of the electrical poles in the frequency spectrum.

All stated values are effective parameters that, within our model, reproduce the behavior of the real-world detectors. However, they are not necessarily a unique combination of parameters that do so. We can therefore not claim that they correspond exactly to the true underlying physical quantities. For this reason, we do not report uncertainties, but only the exact values we used. We expect that large systematic uncertainties exist for our estimates.

\begin{table}[]
    \centering
\begin{tabular}{lccc}
\hline
\textbf{Quantity} & \textbf{Li1P} & \textbf{Li1L} & \textbf{Li2P} \\ \toprule
$V_{f}$ (mm$^3$) & $4.08\cdot10^{-4}$ & $1.44\cdot10^{-5}$ & $4.08\cdot10^{-4}$ \\ 
$V_a$ (mm$^3$) & $4\cdot10^{3}$ & $2\cdot10^{2}$ & $4\cdot10^{3}$ \\ \midrule
$C_e$ (pJ/mK) & 2.11$\cdot10^{-3}$ & 3.5$\cdot10^{-5}$ & 2.5$\cdot10^{-3}$ \\ 
$C_a$ (pJ/mK) & 0.113 & 1.61$\cdot10^{-4}$ & 9.7$\cdot10^{-2}$ \\ \midrule
$G_{eb}$ (pW/mK) & 0.123 & 2.66$\cdot10^{-2}$ & 0.138 \\ 
$G_{ab}$ (pW/mK) & 1.565 & 9$\cdot10^{-3}$ & 1.16 \\ 
$G_{ea}$ (pW/mK) & 0.214 & 2.27$\cdot10^{-3}$ & 0.1 \\ \midrule
$\tau_n$ (s) & 3.82$\cdot10^{-4}$ & 9.4$\cdot10^{-5}$ & 4$\cdot10^{-4}$ \\ 
$\tau_{TP}$ (s) & 4.97$\cdot10^{-3}$ & 2.98$\cdot10^{-3}$ & 4.97$\cdot10^{-3}$ \\ \midrule
$\epsilon$ & 0.115 & 0.056 & 0.104 \\ 
$\delta$ & 0.144 & 0.26 & 0.056 \\ 
$\delta_H$ & 0.9 & 0. & 0.8 \\ \midrule
$R_s$ ($\Omega$) & 4$\cdot10^{-2}$ & 4$\cdot10^{-2}$ & 4$\cdot10^{-2}$ \\ 
$R_H$ ($\Omega$) & 6.75 & 7.25 & 5.25 \\ 
$R_{f0}$ ($\Omega$) & 0.11 & 0.115 & 0.1 \\ \midrule
$L$ (H) & 3.5$\cdot10^{-7}$ & 3.5$\cdot10^{-7}$ & 3.5$\cdot10^{-7}$ \\ \midrule
$T_c$ (mK) & 30.7 & 23.0 & 29.4 \\ 
$k$ (1/mK) & 4.4 & 13.5 & 5.52 \\ \midrule
$\beta$ & 2.25$\cdot10^{-2}$ & 6.25$\cdot10^{-3}$ & 2$\cdot10^{-2}$ \\ 
$I_H$ ($\mu$A) & 4.8 & 0.904 & 8.27 \\ \midrule
$i_{sq}$ (pA/$\sqrt{\text{Hz}}$) & 1.2 & 1.2 & 1.2 \\ 
$E_{J}$ & 1 & 1 & 1 \\ 
$\frac{\Delta R_{f,flicker}}{R_{f0}}$ (pJ) & 8$\cdot10^{-5}$ & 1$\cdot10^{-4}$ & 8$\cdot10^{-5}$ \\ 
$\alpha$ & 2 & 1 & 2 \\ 
$p_0$ & $1.5\cdot10^{-5}$ & $3\cdot10^{-5}$ & $3\cdot10^{-5}$ \\ 
$p_1$ & $1\cdot10^{-5}$ & $2\cdot10^{-5}$ & $2\cdot10^{-5}$ \\ 
$p_2$ & $1.5\cdot10^{-5}$ & $2\cdot10^{-5}$ & $2\cdot10^{-5}$ \\ \midrule
$IB_{min}$ ($\mu$A) & 0.5 & 0.5 & 0.5 \\ 
$IB_{max}$ ($\mu$A) & 17.9 & 17.9 & 17.9 \\ \midrule
$\eta$ (V/$\mu$A) & 5.77 & 5.77 & 5.77 \\ \bottomrule
\end{tabular}
    \caption{Values used in the simulation of the detectors Li1P, Li1L and Li2P. See text, Sec.~\ref{sec:setup} and \ref{app:noise} for definitions.}
    \label{tab:pars_used}
\end{table}

When we randomized the values stated in Tab.~\ref{tab:pars_used} for the detector version used for training in Sec.~\ref{sec:virtual} we multiplied them by a Gaussian random number $\sim \mathcal{N}(\mu=1, \sigma=0.2)$. For every detector version we tested if the transition was within reach with the available heating and bias currents and resampled the parameters with a new random seed otherwise.

When we adjusted the values to two TES versions used in Sec.~\ref{app:multicomponent}, we copied the parameters of the TES to a third thermal component, with an independent readout circuit, but divided the collection efficiency $\epsilon$ by two, assuming that the athermal phonon population is shared equally among the two TES. We also divided the size of the TES and its thermal links to the heat bath and the absorber by two.

\section{Details of models and training} \label{app:training}

We use for the training and evaluation of models the Vienna CLIP computing cluster. For the training and evaluation of neural networks models we used the PyTorch library \cite{NEURIPS2019_9015}. 

\subsection{Details of training in virtual environment} \label{app:training_virtual}

We ran individual single-CPU jobs for the training of each detector version, they took between one and two hours of CPU time to be completed. The detector versions with two TES modules needed between 6 and 8 CPU hours of training time. Depending on the number of gradient steps about half of the time was spent in the simulation which is CPU-bound in our current implementation. We would therefore not gain significantly through operation on a GPU.

The policy and value functions of our SAC agents are 2-layer neural networks with 256 nodes in each layer, and ReLU activation functions.

We optimized their weights with the ADAM optimizer \cite{kingma2017adam}, using a weights decay of $1\cdot10^{-5}$. The learning rates, batch sizes, $\gamma$ values of the temporal difference method, and the number of gradient steps after each environment step were different in each scenario of detector versions and stated in Fig.~\ref{fig:pars_studies}. For training the 2 TES detector versions, we used the same hyperparameters as in the first scenario. We put the $\tau$ update parameter of the SAC algorithm to a value of 0.005 and the initial entropy coefficient $\alpha$ to 0.2. The training was started as soon as one full batch of state transitions was collected in the replay buffer. Gradients were clipped at a maximum norm of 0.5. 

We calculated the target entropy for the policy function as the entropy of a multivariate normally distribution with a given standard deviation. We fixed the number of dimensions of the Gaussian to the size of the action space and the standard deviation at the beginning of the training to the value 0.088. We reduced the target standard deviation of the Gaussian with every gradient step such that it reached 0.05 times its original value at the end of the training.

\subsection{Study of hyperparameters in the virtual environment} \label{app:training_hyperpar}

\begin{figure*}[!t]
\centering
\includegraphics[width=\linewidth]{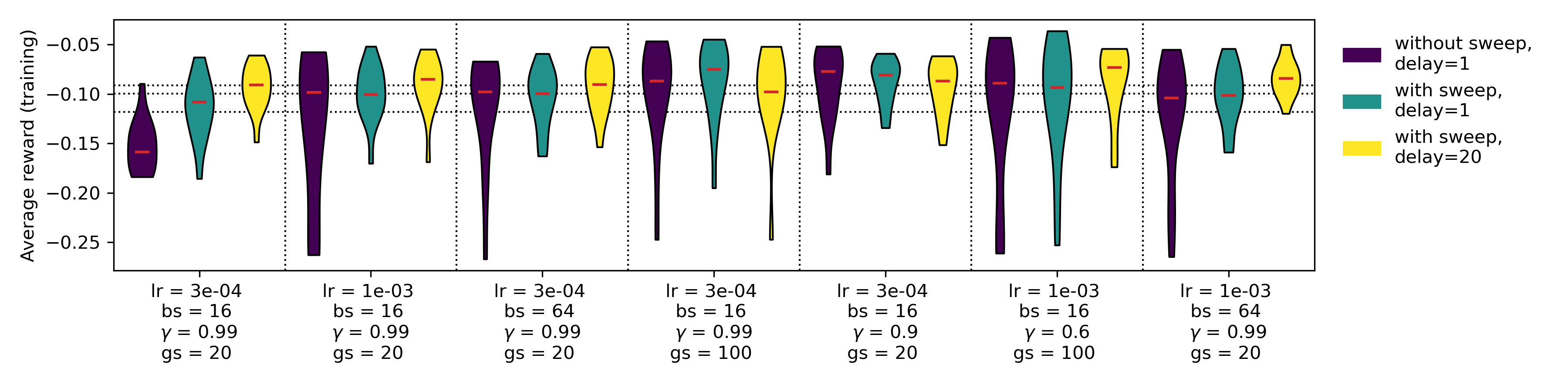}
\caption{Average reward during all training episodes of the different versions of the virtual detectors, grouped into violins for each scenario (violet, turquoise, yellow) and setting of the hyperparameters learning rate (lr), batch size (bs), $\gamma$ and gradient steps (gs). Each violin includes five versions of Li1P, Li1L, and Li2P each, sampled and trained with individual random seeds. The red bars indicate the mean values of the violins and their thickness the density of the represented return distribution. The dotted horizontal lines indicate the mean values of the collected returns of the first (lowest), second, and third (highest) scenarios of detector versions. The dotted vertical lines separate the violins with different hyperparameter settings. The values for the hyperparameters are written in the ticks on the abscissa.}
\label{fig:pars_studies}
\end{figure*}

The results from different hyperparameter settings and scenarios in the virtual environment are visualized in Fig.~\ref{fig:pars_studies}. Several systematic effects are visible. From this study, only the combined effect of faster training and higher obtained final rewards can be accessed. 

First, the initial sweep performed in the second and third scenarios seems to have a positive systematic impact on the obtained rewards. This is expected: without an initial sweep, the agent starts sampling actions in a Monte Carlo style at the beginning of the training, since it has no knowledge of the island in parameter space that leads to an OP in the superconducting transition. This has two disadvantages compared to the grid search that we call sweep. First, we can generalize the computational cost of evaluating a function on a grid compared to evaluating it by Monte Carlo random sampling from the error estimates in the theory of finite elements. From there we know that the computational cost for an evaluation on the grid is lower for spaces with less than 4 dimensions. It is therefore expected, that the grid search will find specific regions in the 2-dimensional action space faster than the Monte Carlo style sampling. Second, in this initial exploration period, the agent performs large jumps between control values. These can lead to thermal relaxations on a longer time scale than the environment steps, and can therefore make the identification of beneficial parameter space more difficult. The situation changes drastically once the agent discovers the islands in parameter space that correspond to the superconducting transition. In this situation the agent learns from every new data point it takes and can converge towards the regions that lead to the highest returns, significantly outperforming any standard Monte Carlo or grid search method. The combination of the initial, fast grid search and the finetuning with the temporal difference method seems therefore to be a good choice for our size of action space. Starting from 4 or more action space dimensions, we expect that the initial grid search would lose its advantage.

Another observation is that a higher number of gradient steps can lead to a higher risk of overall failure. Also this is expected since more gradient steps per data point taken effectively lower the exploration period and can cause the agent to converge towards fluctuations outside the beneficial region of parameter space. A similar effect can be attributed to a higher learning rate and a larger batch size. The results with a discounting factor $\gamma=0.9$ were better compared to the default value $\gamma=0.99$. The value $\gamma=0.6$ obtained a significantly higher risk of failure, however, the effect is hard to disentangle from the effect of the higher number of gradient steps.

The delayed heating did not have a negative impact on the performance during training, on the contrary, it even had a slight positive impact. The reason for that is not ultimately clear. It is e.g.~possible that the agent spends more time during the training period, which is dominated by exploration, in certain regions of the control parameter space and can identify them more easily as opportune or unfavorable. The effect was reproducible in follow-up experiments with different random seeds, we therefore believe that it is not due to random fluctuations.

\subsection{Details of live training} \label{app:training_live}

\begin{table*}[!ht]
    \centering
    \begin{tabular}{lcccccc}
    \toprule
        \textbf{Detector} & \textbf{Li1P} & \textbf{Li1L} & \textbf{Li2P} & \textbf{Li2P} & \textbf{Li2P} & \textbf{Li1L} \\
        \textbf{run} & \textbf{1} & \textbf{1} & \textbf{1} & \textbf{2} & \textbf{3} & \textbf{2} \\
        \midrule
        $\gamma$ & 0.9 & 0.99 & 0.99 & 0.99 & 0.99 & 0.99\\
        Reward & Eq.~(\ref{eq:raw_target}) & Eq.~(\ref{eq:raw_target}) & Eq.~(\ref{eq:raw_target}) & Eq.~(\ref{eq:raw_target}) & Eq.~(\ref{eq:adapted_target}) & Eq.~(\ref{eq:adapted_target})  \\
        TP int. (s) & 20 & 20 & 10 & 10 & 10 & 10 \\
        $DAC_{max}$ & 10 & 10 & 5 & 5 & 5 & 10 \\
        $IB_{min}$ & 0.5 & 0.1 & 0.5 & 0.5 & 0.5 & 0.1 \\
        $IB_{max}$ & 5 & 3 & 5 & 5 & 5 & 3 \\
        ADC range & $\pm10$ & $\pm0.3$ & $\pm1$ & $\pm1$ & $\pm1$ & $\pm0.3$ \\
        TPA in state & yes & no & no & no & no & yes \\
        CPH in state & no & no & no & no & yes & yes \\
        ADCs/$IB$ & no & no & yes & no & yes & yes \\
        $\omega$ & 0 & 0 & 0 & 0 & 0.01 & 0.01 \\
    \bottomrule
    \end{tabular}
    \caption{Hyperparameter and settings of the RL problem used for the six performance runs on the CRESST underground setup. See the text for explanations.}
    \label{tab:live_hyperpars}
\end{table*}

For live training on the CRESST setup, we connected a computing node of the Vienna CLIP cluster with an SSH tunnel to the MQTT broker that was operated in the LNGS network. 

We used for all performance runs the same neural network architectures for policy and value functions as for training in the virtual environment. The neural networks were trained with the ADAM optimizer, with a batch size of 16 and 20 gradient steps, respectively after each test pulse, a learning rate of $3\cdot10^{-4}$ and weight decay of $1\cdot10^{-5}$. Usually performing the 20 gradient steps took much less time than the interval between test pulses, the total time required for training was therefore determined by the measurement time on the experiment. Gradients were clipped at the norm 0.5.

We set the initial entropy of the SAC algorithm to $0.2$, and the $\tau$ update parameter to $5\cdot10^{-3}$. Several hyperparameters and settings of the RL problem were varied throughout the six performance runs, and an overview of them is contained in Tab.~\ref{tab:live_hyperpars}. We fixed the target entropy to the entropy of a 2-dim Gaussian with a standard deviation of 0.088. Contrary to our training in the virtual environment, we did not reduce this value while training progressed.

In some of the runs the reward function was weighted with the inverse $TPA$ values (see Sec.~\ref{sec:Analysis}). The time interval between test pulses was set to a higher value than the default of 10 seconds for two runs, to test if thermal relaxations on larger time scales have an impact on the training. However, no such impact was observed. The normalization intervals for the $DAC$ and $IB$ values were individually adjusted for the detectors, as well as the value range of the analog-digital converter (ADC). The state space of the RL problem was adjusted to contain the $TPA$ and $CPH$ values for a subset of the runs. Furthermore, the division of the values that scale with the ADC and $IB$ ($PH$ and $RMS$) by the $IB$ value is done for a subset of the runs. 

From the limited amount of data, it is hard to make general statements about the impact of certain adjustments on the environment. We can see that the runs that included the $TPA$ value in the state took generally longer to converge, by comparing Li1L runs 1 and 2, and Li2P runs 2 and 3. However, we cannot fully disentangle this effect from the impact of the weighting of the reward function, and the regularization factor $\omega$. Furthermore, by comparing Li2P runs 1 and 2 it seems to us that dividing the ADC-dependent values by $IB$ has a positive impact on the speed of convergence. These observations match our expectations, namely that a simpler state space with fewer redundancies leads to faster learning. However, we cannot exclude that our observations are dominated by random fluctuations.

It is also important to note, that our systematic study of hyperparameters in the virtual environment was not concluded before our time slot for the experiments on the live setup. Therefore not all learnings from the virtual environment could be used for the live runs.

\section{Regularization of jumps in inference} \label{app:regularization}

\begin{figure}[!t]
\centering
\includegraphics[width=\linewidth]{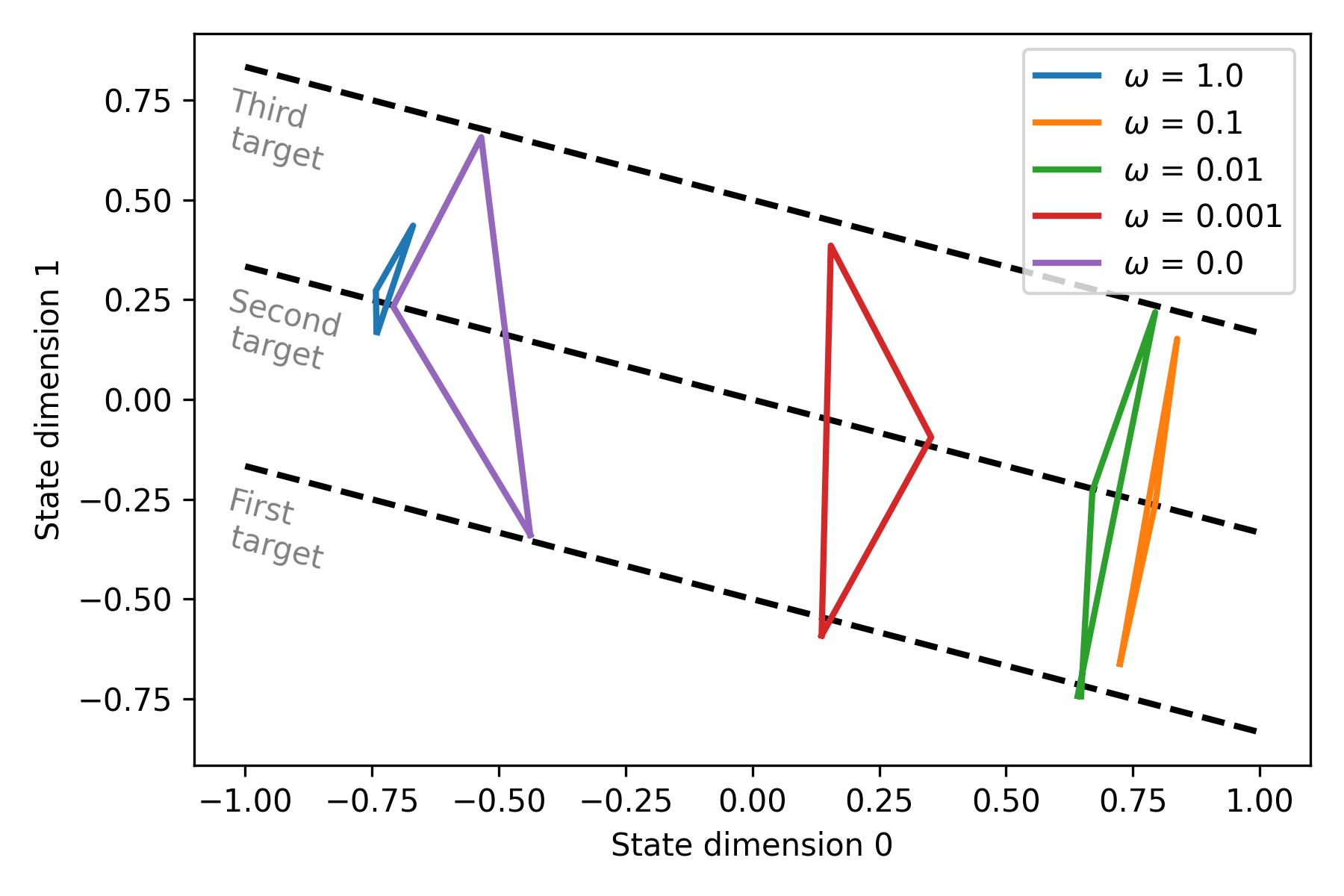}
\caption{Movement of a SAC agent in a two-dimensional toy box environment. The goal of the agent is to jump close to the cyclically changing target lines (black dashed). The paths taken by an agent in inference trajectories are drawn with colored lines, for different magnitudes of the jump regularization parameter $\omega$.}
\label{fig:issue_sac}
\end{figure}

We dedicate this section to studying the impact of the regularization term $-\omega(A_1 - A_0)^2$, which we added to the reward function. Here $A_1$ and $A_0$ are the current and previously taken actions, and therefore the set control parameters. Our intention of this regularization is to mitigate random jumps of the control parameters that are caused by the initial, random initialization of the neural networks, and are not useful for obtaining higher rewards. 

To study this effect we use a toy environment. For this, we imagine the agent to be contained in a two-dimensional box, where certain positions lead to a higher reward than others. With its actions the action jumps to different positions in the box. We define the action space as the position to which the agent wants to jump, therefore a two-dimensional box in the value range -1 to 1. The first two dimensions of the state space are the current position of the agent in the box, and therefore a record of the previously taken action. The state space has a third dimension that cyclically takes the values -1, 0, and 1, with no regard for the actions of the agent. This third dimension of the state space signals the position of a target line in the state space to the agent, i.e.~a line along which the agent will obtain the highest rewards. The position of the target line changes cyclically, and the reward is defined as the Euclidean distance between the position of the agent and the closest point on the target line. This situation is depicted in Fig.~\ref{fig:issue_sac}.

The optimal policy to solve this toy problem is simple: the agent has to find a point on each of the target lines and understand in which order it has to jump between these three points. Which point the agent chooses along the target line does not have any impact on the obtained rewards. The optimal policy is therefore not unique. We trained our SAC agent to learn this policy, with the expected outcome. The trajectory that the agent follows is shown in Fig.~\ref{fig:issue_sac} (violet, $\omega=0$). The position that is chosen by the agent along the target lines is only determined by the random initialization of the network, which leads to convergence to the closest accessible minimum in its loss function that simultaneously satisfies the weight regularization applied by the ADAM optimizer.

We introduce now our manual regularization term that penalizes large jumps of the agent. We train our agent for four choices of the regularization parameter $\omega$: 0.001, 0.01, 0.1, and 1.0. The behavior that we want to enforce is, that the agent should jump between the target lines along the shortest possible path, i.e.~along a line orthogonal to the target lines. We can observe that for the smallest value of $\omega$ (Fig.~\ref{fig:issue_sac}, red) the regularization has practically no impact. In this regime, the effect of the weight regularizer fully dominates over the effect of the jump regularizer that we introduced. For $\omega=0.01$ (Fig.~\ref{fig:issue_sac}, green) and 0.1 (Fig.~\ref{fig:issue_sac}, orange) we can see the desired effect: the agent moves almost orthogonally to the target lines. Good choices for $\omega$ are likely contained in this value range for the toy problem at hand. The choice itself is a trade-off: we can see that for $\omega=0.1$ the effect of the introduced regularization leads to a slight deviation from the target lines, which is generally not the desired outcome. However, the agent's movement is less orthogonal to the target lines for the setting $\omega=0.01$. Finally, for a very large setting of $\omega$, we can see that the agent stays almost at the central target line, and stops responding to the environment. 

\section{Entropy mismatch for action fine-tuning} \label{app:entropy}

\begin{figure}[!t]
\centering
\includegraphics[width=\linewidth]{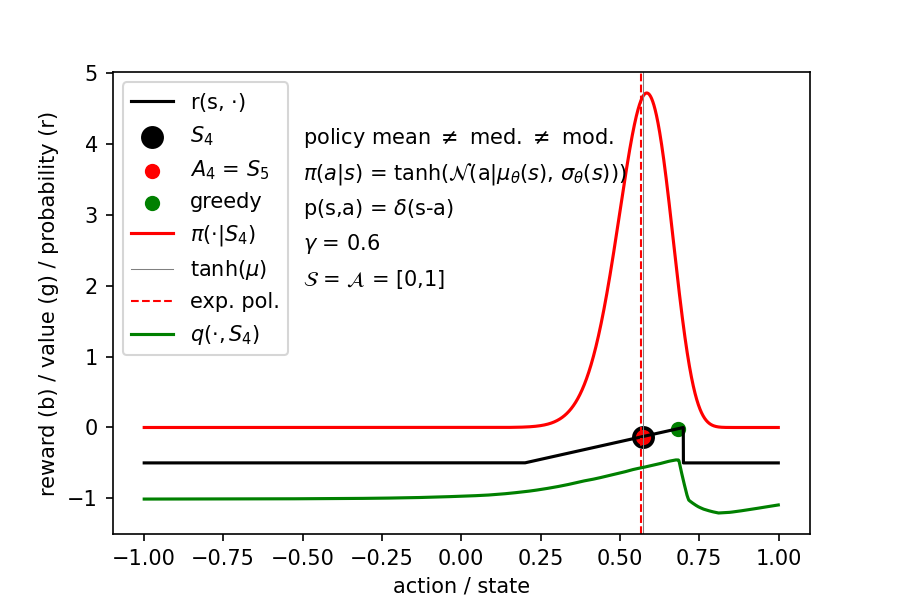}
\caption{Toy environment of an agent that climbs a mountain. The reward function (black) has the shape of a side view of the mountain that ends in a cliff on one side. The critic function (green) learns the shape of the reward function sufficiently well. The policy function (red) learns to overlap with the mountain, instead of placing its expected value on top of the mountain (see text for details). To keep all actions in the interval between -1 and 1, the actions sampled from the Gaussian are again input to a hyperbolic tangent function in the SAC algorithm. This leads to a deviation between the mean, the median, and the mode of the resulting probability distribution. We show by comparing the expected/mean value of the policy (red dashed) with its median (grey) and its modus (peak of the red Gaussian) that this effect does not play a significant role in our experiment. }
\label{fig:issue_sac_entropy}
\end{figure}

The SAC algorithm enforces a target entropy to its Gaussian policy function. This keeps the policy from collapsing to small features in the action space early in the exploratory period of the training, effectively hindering the agent in its exploration. The automated entropy-based tuning of the temperature parameter $\alpha$, which governs the exploration versus exploitation trade-off in the loss functions of the actor and the critic, is one of the key features of the algorithm (see Ref.~\cite{haarnoja2019soft}). For other entropy-based algorithms, an educated guess for this parameter must be made beforehand.

However, the entropy target leads to a side-effect that can be problematic: since the policy function is forced to have a certain width, it has problems resolving structures known to the critic function on smaller scales than its own width. We explain this issue by using a simple toy model, that is also visualized in Fig.~\ref{fig:issue_sac_entropy}. 

We design an environment that corresponds to the side-view of a mountain, which ends with a cliff on one side (Fig.~\ref{fig:issue_sac_entropy}, black line). The reward is equal to the height of the agent on the mountain. The state and action space are both one-dimensional, where the action determines the position in the state space to which the agent jumps next. The optimal policy for this environment is, to jump directly to the peak of the mountain and stay there.

We trained a SAC action on our toy environment. The agent learned to immediately jump in inference trajectories from its starting position to a high position on the mountain, which is the expected behavior. However, the expected value of the policy function, and therefore the action that is chosen in inference trajectories, does not correspond to the peak of the mountain, but to a position displaced towards the flatter mountain slope (Fig.~\ref{fig:issue_sac_entropy}, red Gaussian probability distribution and red dot). This is counter-intuitive since the critic function learned the shape of the reward function very well, and its maximum corresponds almost perfectly to the peak of the mountain (Fig.~\ref{fig:issue_sac_entropy}, green line, and green dot). The reason for this mismatch is, that the SAC algorithm is trained as a stochastic policy, i.e.~it is expected that also during inference trajectories actions are sampled from the probability distribution. The agent therefore seeks to overlap an as large part of the policy distribution with the mountain, instead of shifting the peak of the Gaussian to the peak of the mountain. The entropy-enforced width of the Gaussian leads then to the magnitude of the misplacement between the Gaussian and mountain peaks. 

However, for our purpose, where we want to tune parameters to their optimal values, we want to make deterministic instead of stochastic choices, i.e.~picking the expected value of the policy function in inference trajectories. We resolved this issue for our experiments in the virtual environment by gradually lowering the entropy target throughout the training, which enables the SAC agent to converge towards finer structures in the control parameter space. 

\section{Detector designs with more components} \label{app:multicomponent}

It is straightforward to generalize the ODE system describing a cryogenic detector to designs with more thermal components or TES. Such devices were operated in previous CRESST runs, e.g.~composite designs with a separate carrier crystal \cite{angloher_results_2016} and mounting structures instrumented with individual TES \cite{PhysRevD.100.102002}. Designs with TES separated on a remote wafer were proposed for the operation of delicate target materials \cite{angloher_first_2023}. For such non-standard scenarios, the electrothermal system can be jointly written in matrix-vector notation:
\begin{align}
    \underline{\dot{T}}(t)&=\operatorname{diag}(\underline{C})^{-1}\biggl(\underline{P}\left(t, \underline{T}(t), \underline{I_{f}}(t)\right) \nonumber \\
    &+\operatorname{diag}\left(\underline{G_{b}}\right)\left(\underline{T_{b}}-\underline{T}(t)\right)+(\underline{\underline{G}}-\operatorname{diag}(\underline{\underline{G} 1})) \underline{T}(t)\biggr), \\
    \underline{\dot{I}_{f}}(t)&=\operatorname{diag}(\underline{L})^{-1}\biggl(\operatorname{diag}\left(\underline{R_{s}}\right) \underline{I_{b}} \nonumber \\
    &-\operatorname{diag}\left(\underline{I_{f}}(t)\right)\left(\underline{R_{f}}(\underline{T}(t))+\underline{R_{s}}\right)\biggr),
\end{align}
\noindent where underlined (double underlined) quantities are vector (matrix) valued, and $\underline{\underline{G}}$ describes the symmetric matrix of thermal couplings between components. All other quantities are equivalently generalized to vectors from Eq.~(\ref{eq:odes}). As we generally neglect noise contributions that do not originate in the TES or the readout circuit directly, our previously introduced description of the detector noise remains unchanged. Designs with multiple TES and heaters are typically harder to optimize as the problem's dimensionality grows with the number of control parameters and observables to optimize. To show the applicability of our method also for this more challenging regime we adjust for our studies in Sec.~\ref{app:multicomponent} the parameters of Li1P, Li1P, and Li2P such that they resemble a scenario for cryogenic detectors with two TES and correlated heatings.

\subsection{Training and results} \label{app:results_multicomp}

\begin{figure}[!b]
\centering
\includegraphics[width=\linewidth]{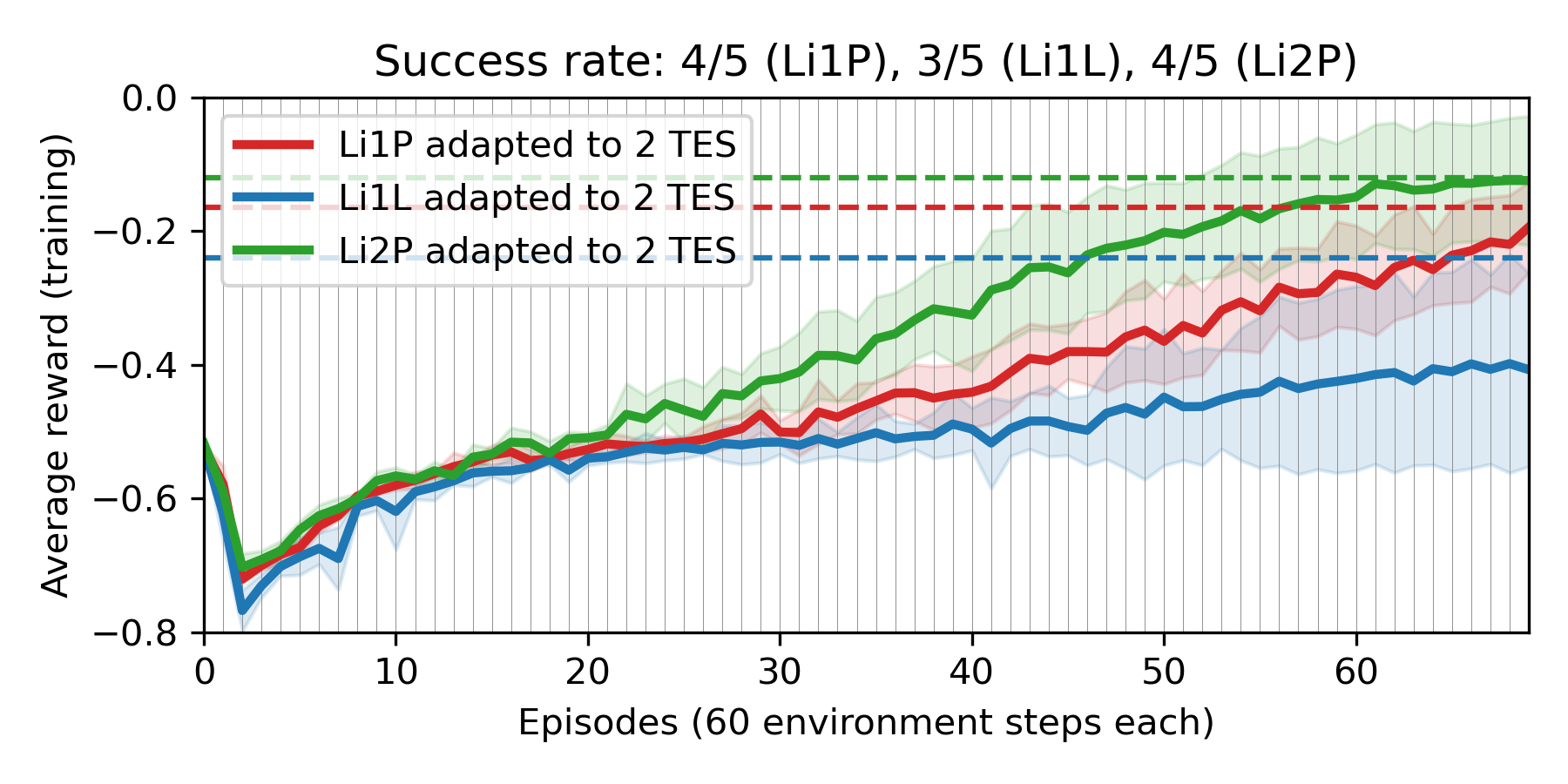}
\caption{Return per episode for Li1P (red), Li1L (blue) and Li2P (green) adjusted to two TES. The thick line represents the mean of five trained versions of the detectors, sampled with different random seeds. The shaded region shows the upper and lower standard deviations. We show benchmarks (dashed lines) for all three detectors. These benchmarks were calculated by taking the average reward in the last episode of the training for all versions of the single-TES detectors that were trained in Sec.~\ref{sec:virtual}, and multiplying it by two. The benchmark is reached by Li1P and Li2P, but not by Li1L.}
\label{fig:rewards_virtual_double}
\end{figure}

We investigated how the measurement time required for training scales with the dimensionality of the control parameter space, and the complexity of the detector design. We apply the equivalent procedure described throughout Sec.~\ref{sec:virtual} to our adaptions of Li1P, Li1L, and Li2P with two TES. We make no adaptations to the default training procedure (first scenario) and the default (first) set of hyperparameters. The training is repeated for five versions and random seeds and the resulting average rewards during training are visualized in Fig.~\ref{fig:rewards_virtual_double} for the detectors separately. The training is successful for 4/5 versions of Li1P and Li2P, and 3/5 of Li1L. For the other versions, only one TES transition is found, and the agent sticks to this local return maximum instead of exploring the environment further. In a practical application, this trapping in local optima can be prevented by enforcing more exploration and accepting longer training times. Finding optimal OPs takes roughly twice as long as for the version with a single TES. Since this more complex detector scenario uses 4 instead of the 2 actions as in the standard detector designs, and requires roughly twice the training time, we can hardly extrapolate a scaling behavior for even larger state spaces. In a worst case scenario, the scaling behavior could be exponential, however, since neural networks do not suffer from this curse of dimensionality (exponential scaling behavior) in many other tasks, we have reason to believe that the scaling behavior would be better. In our experiments we used with the two-TES detectors a five times higher target entropy than in the experiments with the standard design. The target entropy is lowered while training progresses. 

\bibliographystyle{unsrtabbrv}
\bibliography{main}

\end{document}